\documentclass[%
 reprint,
superscriptaddress,
nofootinbib,nobibnotes,
amsmath,amssymb,aps,
]{revtex4-1}

\usepackage{siunitx}
\usepackage{graphicx}
\usepackage{dcolumn}
\usepackage{bm}
\usepackage{epstopdf}
\usepackage{bbm}
\usepackage{amsmath,textgreek}

\usepackage[hidelinks]{hyperref}

\makeatletter
\newcommand{\mypm}{\mathbin{\mathpalette\@mypm\relax}}
\newcommand{\@mypm}[2]{\ooalign{%
		\raisebox{.1\height}{$#1+$}\cr
		\smash{\raisebox{-.6\height}{$#1-$}}\cr}}
\makeatother

\makeatletter
\newcommand*{\citenumns}[2][]{%
	\begingroup
	\let\NAT@mbox=\mbox
	\let\@cite\NAT@citenum
	\let\NAT@space\NAT@spacechar
	\let\NAT@super@kern\relax
	\renewcommand\NAT@open{}%
	\renewcommand\NAT@close{}%
	\cite[#1]{#2}%
	\endgroup
}
\makeatother

\usepackage{commath}
\usepackage{tabularx}

\usepackage{booktabs}

\begin{document}

\preprint{APS/123-QED}

\title{Angular momentum anisotropy of Dirac carriers: A new twist in graphene}

\author{M. Prada}%
\email{mprada@physnet.uni-hamburg.de}
\affiliation{%
	I. Institute for Theoretical Physics, Universit\"at  Hamburg HARBOR, Geb. 610 
Luruper Chaussee 149, D-22761 Hamburg, Germany\\
}

\date{\today}

\begin{abstract}
Dirac carriers in graphene are commonly characterized by a pseudospin degree of freedom, arising from 
the degeneracy of the two inequivalent sublattices.  
The inherent chirality of the quasiparticles leads to a topologically non-trivial band structure,
where the in-plane components of sublattice spin and momentum are intertwined. Equivalently, 
sublattice imbalance is intimately connected with angular momentum, 
inducing a torque of opposite sign at each Dirac point.
In this work we develop an intuitive picture that associates sublattice spin and winding number with angular momentum. 
We develop a microscopic perturbative model to obtain the finite angular momentum
 contributions along the main crystallographic directions. 
Our results can be employed to determine the angular dependence of the $g$ factor and 
of light absorption in honeycomb bipartite structures. 
\end{abstract}

\maketitle
\section{Introduction}
\label{intro}
The linear dispersion at the Fermi level of graphene is cognate  with the Dirac cones of 
massless relativistic particles \cite{novoselov2005two}, 
motivating extensive research  towards the parallelism with relativistic quantum mechanics and 
quantum electrodynamics (QED) in  solid-state materials 
\cite{katsnelson2006chiral,geim2010rise,Nair08,katsnelson2007,GIULIANI2012461,PhysRevX.5.011040,golub}.
In this Dirac-type model, the notion of sublattice (SL) spin  takes the role of real spin, with  \textquotedblleft up" 
and \textquotedblleft down" states being associated with the two SL components that constitute the honeycomb structure. 
The inherent chirality  of the  Dirac carriers leads to  a topologically non-trivial band structure,
where the related edge's helicity gives rise to  the spin Hall insulator \cite{kane2005quantum,kane2005z,jonas}.  
The presence of a pseudospin degree of freedom in the Hamiltonian allows a parallelism between
its torque and that of the angular momentum \cite{regan,chino}, where 
an experimental connection has been realized only in photonic graphene \cite{Song15,Liu2018,Liu2020}.
 
In this work  we evaluate 
the constants of motion and provide an intuitive connection between SL spin and angular momentum
in the neighborhood of the Fermi energy. 
We develop a microscopic perturbative model to obtain the corrections to the angular momentum 
$\vec L$ in terms of atomic parameters. We consider band hybridization, spin-orbit coupling (SOC) 
and Bychkov-Rashba effect, and 
obtain a peculiar anisotropy in the corrections, owing to the relation of $\hat L$ and SL spin. 
These corrections have been verified in large graphene samples, by measuring the 
\textit{g}-factor corrections \cite{exp}. 

This paper is organized as follows: First we employ symmetry arguments to analyze the low-lying
quasiparticle Hamiltonian. 
We examine perturbatively the $d$-orbital contribution to the  $\pi$-bands ($l=2$,  $m_l = \pm 1$), 
and calculate the correction to the angular momentum. 
Finally, we include perturbatively the mixing with the $\sigma$-band ($l=1$,  $m_l = \pm 1$) 
via atomic SOC and Bychkov-Rashba interaction, and obtain the 
angular momentum corrections in terms of  atomic parameters.

\section{Band mixing and hybridization}
\subsection{Dirac electrons to lowest order: $\langle \hat L\rangle =0$ orbitals}
\label{pz}
Dirac electrons near the Fermi edge are commonly described by $\pi$ orbitals within 
a nearest neighbors (NN) tight-binding model \cite{neto2009electronic, katsnelson2012graphene}.
The two sublattices that constitute the honeycomb structure lead to two energy bands,
whose interplay yields a conical quasiparticle spectrum. 
The concept of isospin or SL 
spin is commonly introduced, where the $z$ component, $\sigma_z$, accounts for the SL occupation imbalance. 
At the Dirac points (DPs), 
the quasiparticle Dirac-type Hamiltonian reads as \cite{kane2005quantum,regan,neto2009electronic,katsnelson2012graphene}:
\begin{equation} 
\label{eqH0} 
\hat H_0 = \hbar v_F \left(\tau q_x\hat \sigma_x   +  q_y\hat \sigma_y \right)+ 
\hat \lambda \hat \sigma_z
= \hbar v_F  \hat{\vec\sigma}_\tau^\parallel  \hat{\vec q} 
+\hat\lambda\hat \sigma_z , 
\end{equation} 
where we have assigned the valley index $\tau$ = 1 and $\tau$ = -1, respectively, to the DPs, $K$ and $K^\prime$,
and  $\vec  q = (q_x,q_y)$ is the small vector off the nearest DP.
Here, $v_F\simeq 10^6$ ms$^{-1}$ the Fermi velocity, 
$\hat{s}_\alpha$, $\hat{\sigma}_\alpha $ are the Pauli matrices representing the electron spin and sublattice-spin, respectively, and  
we have defined the vector $ \hat{\vec\sigma}_\tau^\parallel  \equiv (\tau \hat\sigma_x, \hat\sigma_y)$. 
The first term originates from the quantum mechanical hopping between the two sublattices and 
tends to align  $\vec  q$ and $ \hat{\vec\sigma}_\tau^\parallel$.  
In spite of involving the Pauli matrices, it is a {\it scalar} (see Appendix \ref{appa}).
The second term can be, however, a scalar, in the case of a  Kane-Mele SOC \cite{kane2005quantum} 
($\hat\lambda\hat\sigma_z = \tau\lambda_I\hat s_z\hat\sigma_z $) 
or a pseudoscalar, in case of a staggered sublattice potential \cite{semenov}, 
$\hat\lambda\hat\sigma_z = \varepsilon_{AB}\hat\sigma_z =(\varepsilon_A-\varepsilon_B)\hat\sigma_z/2$, owing to the broken parity symmetry.
The distinction is important, as it is only in the former case where a direct analogy 
with the equations of a dipole in a magnetic field can be made. 
To lowest order, the eigenenergies of (\ref{eqH0}) are given by: 
\begin{equation}
\varepsilon_\pm = \pm\sqrt{(\hbar v_F |q|)^2 +  \lambda^2}. 
\label{eqGS}
\end{equation}
The corresponding eigenstates are commonly given in terms of the dominant ($p_z$-orbital) contribution at 
sublattices $A$ and $B$ for $q\neq 0$, 
\begin{equation}
|\varphi^{(0)}_\pm \rangle \simeq 
c_A|p_z^A\rangle 
+ c_B|p_z^B\rangle, \ \frac{c_B}{c_A} = \pm e^{i\tau\varphi_q } , 
\ \varphi_q= \arctan{\frac{q_y}{q_x}}.
\label{eq0}
\end{equation}
The relative phase between the two sublattice components $\tau\varphi_q$ results indeed in a Berry phase \cite{shapere},
and determines the  direction of the associated SL spin \cite{ando1998,mceuen1999}, in analogy
with chirality and  charge conjugation in QED \cite{katsnelson2007,katsnelson2006chiral}.
At the critical point $q =0$, however, the SL spin components are decoupled, and the solutions 
depend on the (pseudo) scalar nature of the mass term. A scalar, Kane-Mele SOC results in decoupled bispinors 
where the conduction band in $K$ relates to the valence band in $K^\prime$, leading to the spin-Hall effect \cite{kane2005z,kane2005quantum}.
On the other hand, the  staggered potential results in a trivial insulator, with the peculiarity 
of {\it pseudoscalar} eigenvalues at the DPs. 

We evaluate now the expectation value of the SL spin components.
For the states of (\ref{eq0}), we obtain: 
\begin{eqnarray}
 \sigma^0_z &\equiv&\langle\varphi^{(0)}_\pm |\hat \sigma_z|\varphi^{(0)}_\pm \rangle =|c_A|^2- |c_B|^2, 
\nonumber \\
\langle \hat{\vec\sigma}_\tau^\parallel\rangle^0 &\equiv&\left(
\begin{array}{c}  \langle \tau \hat \sigma_x\rangle\\ \langle \hat \sigma_y\rangle 
\end{array} \right)\simeq 
\pm \left( 
\begin{array}{c}  \tau\cos{\varphi_q\tau}\\ \sin{\varphi_q\tau} 
\end{array}
\right)=\pm \frac{\tau}{q} 
\left( 
\begin{array}{c}  q_x\\ q_y
\end{array}
\right),\nonumber\\
\label{eqq}
\end{eqnarray}
where we have used $2|c_A|^2\simeq 1$.  
A three-component axial vector $\vec\sigma_\tau$ was already identified  
with the real angular momentum \cite{regan, chino}. Doing so, however, results in unphysical 
interpretations for the equations of motion or the conserved quantities, owing to
the different physical meaning  of the different components.  
Here we aim at bringing clarity and physical meaning to the connection between 
the angular momentum components $\hat L_\parallel, \hat L_z$, and  $\vec\sigma_\tau^\parallel$, $\sigma_z$.   
For the easy axis,  
the $z$ component of angular momentum  is associated with the pseudoscalar  $\tau \sigma_z$: 
\[
\left[ \hat H_0, \hat {L}_z\right] = -\frac{\hbar}{2}\left[ \hat H_0, \hat {\sigma}_{z\tau}\right] = -i\hbar
v_F \hat {\vec \sigma}_\tau^\parallel \times \hat{\vec p} = -i\hbar \dot{\vec r} \times \hat{\vec p} ,
\]
where we have used that $\dot{\vec{r}} = (i/\hbar) [ \hat H_0, \hat {\vec r}] =  v_F \hat {\vec \sigma}_\tau^\parallel$, 
that is, $ \hat {\vec \sigma}_\tau^\parallel$ is aligned with the velocity \cite{chino}.  
As Mecklenburg {\it et al.} already pointed out \cite{regan},   
this implies that the constant of motion is $ 2{L}_z +  \hbar \tau{ \sigma}_z$, which we 
generalize to bilayer graphene's Dirac Hamiltonians in terms of the 
winding number $\nu$ (see Appendix \ref{app1b}) as:
\begin{equation}
\frac{\rm d}{\rm d t} \left(2 {L}_z +  \nu\hbar \tau{ \sigma}_z\right) = 0. 
\label{eqLz}
\end{equation}
That means that {\it  sublattice imbalance induces a torque in the quasiparticle spectrum 
of opposite sign on each valley.} Likewise, a finite $L_z$ induces a sublattice imbalance of opposite sign on each valley.
The same line of argument can be used in the presence of a Kane-Mele term, where $\langle\tau\sigma_z\rangle$  has
different sign for electrons with opposite spin, 
resulting in splitting of  left- or right-propagating states \cite{hasan2010colloquium,kane2005quantum}. 
This peculiarity is reflected in the band mixing, as will be detailed in the next section: 
Dirac electrons on a given SL (say $A$) and valley
couple only to clockwise rotating orbitals, and the converse is true for those on SL $B$. 

An intuitive picture can be given in terms of the winding number: in monolayer (bilayer) graphene, 
the Berry phase is $\pi$ (2$\pi$), yielding a $\nu=$1 (2) winding number. 
This means that an adiabatic closed contour in momentum space 
the SL spin \textquotedblleft winds \textquotedblright 
around the origin once (twice), just like  momentum does \cite{asboth,eprada} [see Fig. \ref{figSrot}(c)]. 
This rotational in momentum space can be then intuitively associated with a torque and, hence, its path integral should 
be related to the angular momentum. 


For the in-plane components  we may define an in-plane pseudovector operator 
\begin{equation}
\hat {\vec{\mathcal L}}_\parallel = (\hat \lambda /\hbar v_F)\hat {\vec r} \times \hat{\vec u}_{z} , 
\label{eqPLpar}
\end{equation}
that could 
relate to in-plane angular momentum. 
With this definition, we obtain: 
\begin{equation}
\frac{\rm d}{{\rm d} t}\hat {\vec {\mathcal L}}_\parallel =
\frac{i}{\hbar}\left[ \hat H_0, \hat {\vec {\mathcal L}}_\parallel\right] =
 -\hat\lambda 
\left(\begin{array}{c}-\hat\sigma_y\\ \tau \hat \sigma_x\end{array}\right) = 
\hat \lambda \hat{\vec \sigma}_\tau^\parallel\times\hat{\vec u}_z.
\label{eqLpar}
\end{equation}
On the other hand, we may calculate  $ \dot{\vec\sigma}_\tau^\parallel$ as: 
\begin{eqnarray}
\frac{1}{v_F}\frac{\rm d}{{\rm d} t}\dot{\vec{r}} &=& 
\frac{\rm d}{{\rm d} t}\hat{\vec{\sigma}}_\tau^\parallel = 
\frac{i}{\hbar} \left[ \hat H_0, \hat {\vec {\sigma}}_\tau^\parallel \right]  \nonumber \\&=&   
-2 v_F \tau
\left(\begin{array}{c}-\hat q_y\\ \hat q_x\end{array}\right)\hat\sigma_z  + \frac{\hat\lambda}{\hbar}
\left(\begin{array}{c}-\hat\sigma_y\\ \tau \hat \sigma_x\end{array}\right), 
\end{eqnarray} 
that is, together with (\ref{eqLpar}), we obtain the in-plane relation, 
\begin{equation}
\frac{\rm d}{{\rm d} t}  \left( \hat {\vec {\mathcal L}}_\parallel + \frac{\hbar}{2}\hat{\vec{\sigma}}_\tau^\parallel \right) = 
\hbar v_F \tau\left(\begin{array}{c}-\hat q_y\\ \hat q_x\end{array}\right)\hat \sigma_z  .
\label{eqlpar}
\end{equation}
This implies that 
$2\hat {\vec {\mathcal L}}_\parallel + \hbar\hat{\vec{\sigma}}_\tau^\parallel$ is a constant of motion  as long as SL symmetry 
is preserved, $\sigma^0_z =0$ or strictly at the DPs, if the Hamiltonian contains a non-zero mass term. 
To lowest order, using (\ref{eqq})  in (\ref{eqlpar}), 
we obtain:  
\[
\frac{{\rm d}  {\hat{\vec {\mathcal L}}}_\parallel }{{\rm d} t} \left( 1 -\hbar v_F q \hat\lambda^{-1}\hat\sigma_z \right)
 + \frac{\hbar}{2} 
\frac{\rm d  {\hat{\vec \sigma}}_\tau^\parallel}{{\rm d} t} = 0 , 
\]
Note that the second  term  in the brackets evaluates to zero in the  absence of a mass term, $\lambda =0$. 
Although $\hat \sigma_z$ can be associated with 
$\hat L_z$, 
no connection exists for the in-plane counterparts, however, we can relate 
$\vec \sigma_\tau^\parallel$ to $\hat {\vec {\mathcal L}}_\parallel$ at DPs, which is beyond the scope of 
this work. 

In what follows, we focus on the different angular momentum components and their relation to sublattice imbalance. 
We note that neither a staggered potential  nor a Kane-Mele term provides a valley-sublattice imbalance, as $\langle\tau\sigma_z \rangle =0$
when averaging over the low-lying energy states. Indeed,  the axial symmetry of the $p_z$ orbitals, with $l=1$ and $m_l =0$, 
involves $\langle \hat L_\alpha\rangle =0$, $\alpha = x,y,z$. 
However,  
(i) band hybridization and band mixing, 
(ii) atomic spin-orbit coupling, 
(iii) Bychkov-Rashba effect, and 
(iv) structural spin-orbit coupling 
may finite contributions to the angular momentum. We consider in the following sections 
these general  mixing mechanisms from a microscopic perspective, and provide an intuitive connection 
with SL-spin degree of freedom. 

\subsection{The hybridized $\pi$ bands}
\label{pi}
We consider the $\pi$-band hybridization near the DPs, which is, to lowest order, given by 
$p_z$ orbitals of Eq. (\ref{eq0}).   
Using perturbation theory within the two-center Slater-Koster approximation,   
the contributions from the $d$ band are obtained via hopping to $d_{yz}$ and $d_{xz}$  orbitals,
\cite{konschuh2010tight, huertas2006spin} (see also Appendix \ref{appb}): 
\begin{equation}
|\varphi^{(1)}_d\rangle =  -
\frac{3i\tau V_{pd\pi}}{\sqrt{2}\varepsilon_{pd}}
\left(c_A  |2 \tau\rangle^B + c_B|2 -\tau\rangle^A\right), 
\label{eqdcorr}
\end{equation}
where $\varepsilon_{pd}$ is the energy of the $d$  relative to the $p$ orbitals, with $ \varepsilon_d\gg\varepsilon_p$, 
justifying our perturbative approach and we have employed the angular momentum representation, 
$|2 \pm 1 \rangle =(| d_{xz}\rangle \mp i|d_{yz}\rangle)/\sqrt{2} $. 
It is straightforward to see that the expectation value of the in-plane angular momentum is still zero,  
$\langle \hat L_x\rangle_\pi = \langle \hat L_y\rangle_\pi =0$. 
 We note, however, that within our NN model, the $p_z$ orbital in $A$ sublattice couples to 
$|2 \tau\rangle^B$, which is a clockwise  rotating 
$B$ for $\tau =1$ ($K$ point)  and to the anti-clockwise rotating $B$  in $K^\prime$, 
allowing us to associate SL spin and $L_z$ in an 
intuitive manner. Using (\ref{eqdcorr}), we have: 
\begin{equation}
\langle \hat L_z\rangle_\pi = \langle \varphi^{(1)}_d |  \hat L_z |\varphi^{(1)}_d\rangle = 
\left(\frac{3V_{pd\pi}}{\sqrt{2} \varepsilon_{pd}}\right)^2  \tau \sigma_z^0 \sim \frac{\lambda_I}{\lambda_{\rm soc}^d}\tau \sigma_z^0, 
\label{eqLzPi} 
\end{equation} 
where in the last step we have used the result of Konschuh {\it et al.}\cite{konschuh2010tight}. 
\begin{figure}[!hbt]
\includegraphics[angle=0, width=.85\linewidth]{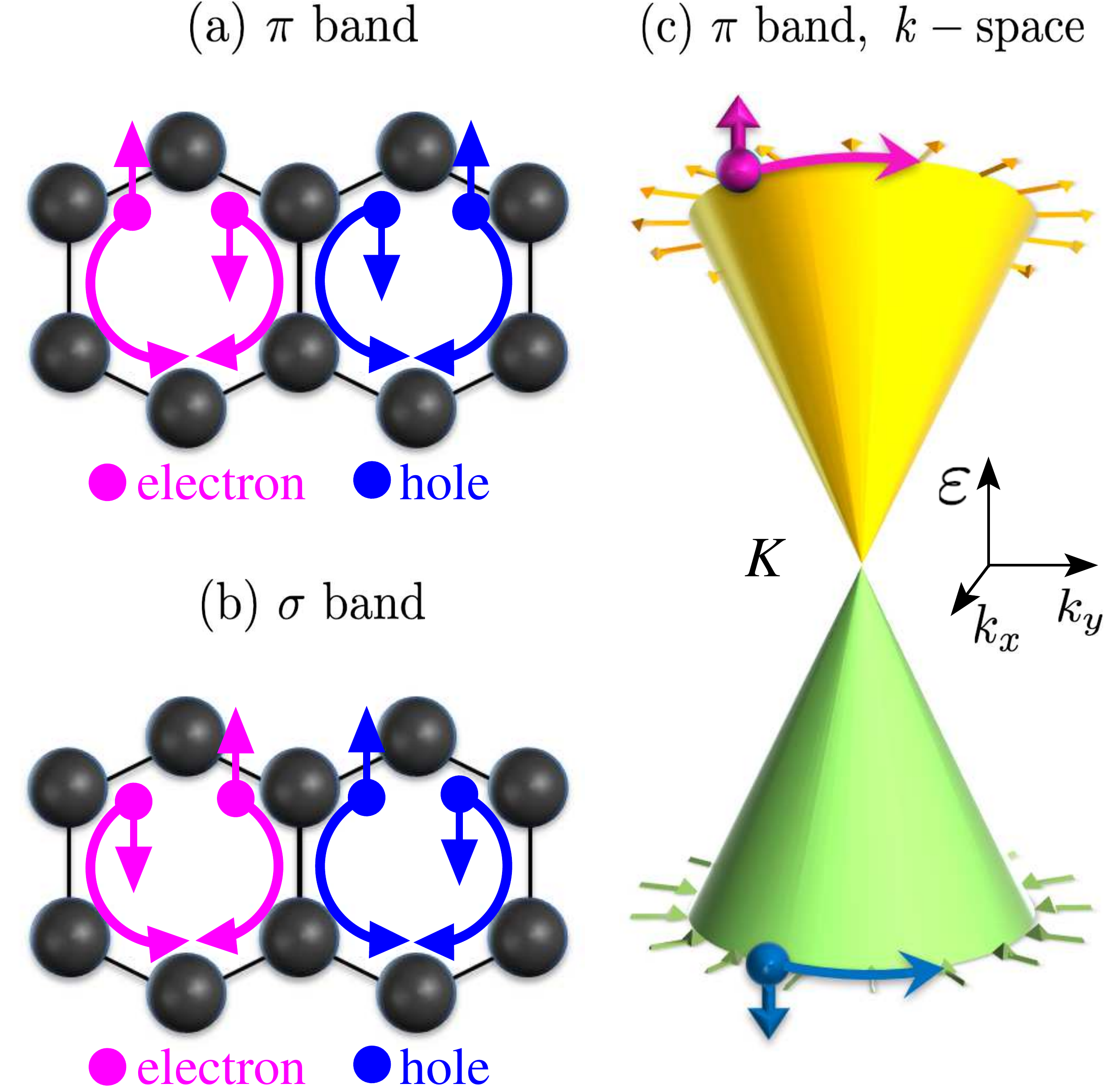}\\
\caption{
Illustration of the spin-valley-orbit coupling of Dirac carriers.
(a) Spin `up' $p_z$ electrons  couple to counter-clockwise ($m_l=1$ or $\tau \sigma_0>0$) rotating $d$-orbitals, 
whereas spin `down' electrons couple to clockwise  rotating ones (magenta). 
The converse occurs for the carriers in the valence band or holes (blue). 
(b) Spin `up' (`down') $p_z$ electrons couple to (counter)-clockwise in-plane rotating $\sigma$-orbitals, 
whereas  spin `down' (`up') holes couple to (counter)-clockwise in-plane $p$-orbitals of the $\sigma$ band. 
(c) Illustration of the CPT symmetry near the Dirac point: a left-handed quasiparticle with positive energy
(magenta) is related to a right-handed one with negative energy (blue). The yellow (green) arrows 
indicate the sublattice spin direction along the iso-energetic paths. 
}
\label{figSrot}
\end{figure}
Figure \ref{figSrot}(a) illustrates, in essence, the spin-valley-orbit coupling for the low-lying energy bands 
showing  charge-conjugation, parity, time-reversal (CPT) symmetry. 
The conduction or electron band (CB) is characterized by $\langle\tau s_z \sigma_z\rangle >0$. That is,  a spin-
\textquotedblleft up\textquotedblright (-\textquotedblleft down\textquotedblright) 
electron has  $\tau\sigma_z > 0$ ($\tau\sigma_z <0$), and hence, acquires a positive (negative)  
correction to the angular momentum (\ref{eqdcorr}), meaning a coupling to a counter-clockwise (clockwise) rotating $d$ orbital. 
That is, the $z$ component of the spin is {\it aligned} with the angular momentum correction. 
For the valence or hole band (VB), however, the converse occurs: having  $\tau s_z \sigma_z = -1$, 
the spin anti-aligns with the angular momentum correction. The system is is thus CPT symmetric: For any left-handed electron 
state with positive energy $\varepsilon_+$, a corresponding conjugated right-handed hole state with 
energy $\varepsilon_-= -\varepsilon_+$ and opposite angular momentum can be found.
The CPT symmetry and,  ultimately, the chirality of Dirac electrons, is reflected in the proportionality to  
$\tau \sigma_z^0$ of the angular momentum corrections [see Fig. \ref{figSrot}(c)].  
\subsection{The hybridized $\sigma$ bands}
\label{sigma}
\begin{figure}[!hbt]
{
\includegraphics[angle=0, width=1.0\linewidth]{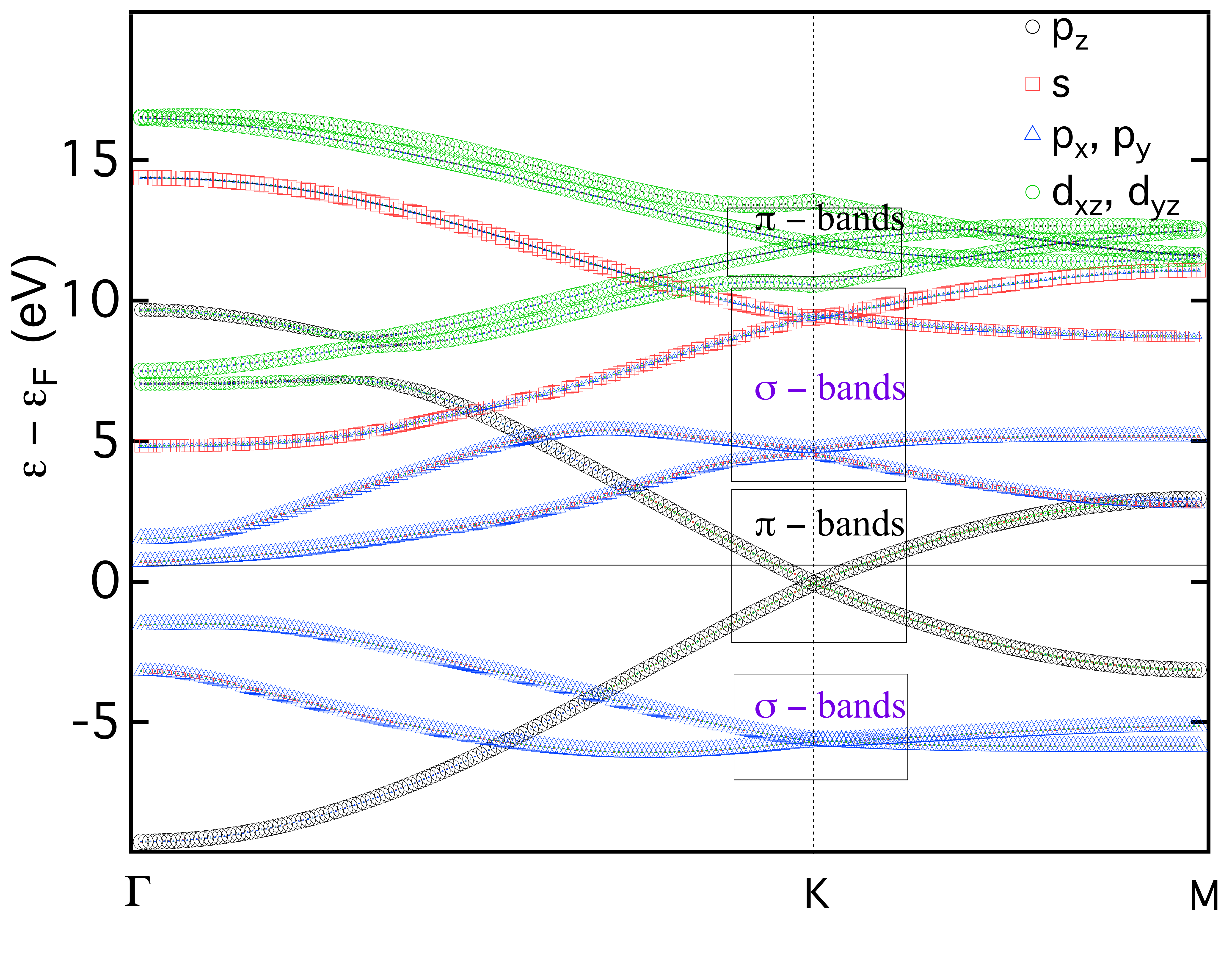}}\\
\caption{(Color online) Calculated bands structure of graphene obtained from a tight-binding model. 
The squares indicate the different hybridizations near the $K$ point, for $\sigma$ or $\pi$ bands.
}
\label{FigBands}
\end{figure}

We consider next the hybridization of the 
$\sigma$ bands near the DPs. 
Once again, we employ the two-center Slater-Koster approximation for an $s$-orbital 
with $p_{x,y}$ on its NN, at $\vec k  = \tau \vec K +\vec q$. 
Employing the angular momentum representation, $|l, m_l\rangle$, and noting that
$|p_x\rangle \pm i|p_y\rangle = \sqrt{2} |1 \mp 1\rangle$, the 
hybridized doublets (see Appendix \ref{appc}) are: 
\begin{eqnarray}
|\phi^+_{AB}\rangle &= &i\tau\cos{\gamma} |s^A\rangle + \sin{\gamma} | 1 \tau \rangle^B,  \nonumber \\ 
 |\phi^+_{BA}\rangle &=& i\tau\cos{\gamma} |s^B\rangle + \sin{\gamma} | 1 -\tau \rangle^A \nonumber \\
 |\phi^-_{AB}\rangle &=& -i\tau\sin{\gamma} |s^A\rangle + \cos{\gamma} | 1 \tau \rangle^B, \nonumber \\ 
|\phi^-_{BA}\rangle &=& -i\tau\sin{\gamma} |s^B\rangle + \cos{\gamma} | 1 -\tau \rangle^A,  \nonumber \\ 
|\phi^0_{AB}\rangle &=&  | 1 -\tau \rangle^B,\quad |\phi^0_{BA}\rangle = | 1 \tau \rangle^A, 
\label{eqS0}
\end{eqnarray}
with corresponding eigenenergies:  
\[
 \varepsilon_\sigma^\pm =\frac{\varepsilon_s}{2}\pm \sqrt{\frac{\varepsilon_s^2}{4} +\frac{9V_{sp\sigma}^2}{2}}, 
\ \varepsilon_0 =0.
\]
Figure \ref{FigBands} illustrates the hybridization of the bands at $K$ point, obtained from a 12-bands tight-binding model. 
The size of the symbols reflects the contribution of the orbitals to the corresponding eigenstates, whereas the colors
represent the different orbitals considered in the model: $s$, $p_{x,y}$, $p_z$, and $d_{xz,yz}$. 
Note that the splittings due to SOC are not visible at this energy scales. 

The $p_z$ orbitals described in (\ref{eq0}) can mix with the $\sigma$ bands 
either  intrinsically  via atomic spin-orbit interaction or 
extrinsically, via, {\it e.g.} Bychkov-Rashba effect \cite{konschuh2010tight,min2006intrinsic, yao2007spin,
rashba2009graphene,gmitra2009band, huertas2006spin,kochan}. 
The latter is linear in uniaxial field, 
and leads to the Stark effect,  consisting of an atomic dipole moment
induced by a perpendicular effective field, $\vec E = E \vec{u_z}$.
Microscopically, the induced dipole  results in  a non-zero intra-atomic coupling between  the $p_z$- and $s$ orbitals.
The intrinsic SOC, on the other hand, couples the $p_z$-orbitals of the $\pi$ bands with the $p_{x,y}$-orbitals of the $\sigma$ bands. 

The $\sigma$ band mixing near the Fermi energy is expected to be smaller than the $d$ band contribution, since 
the spin-orbit coupling parameter and the Stark parameter   $\lambda_z = e E \langle s |\hat z|p_z\rangle $ 
are small compared to the $p$-$d$ coupling, $\lambda_{\rm soc}^p , \lambda_z\ll V_{pd\pi}$. 
We thus consider the $\sigma$ band mixing perturbatively in the following. 

\subsubsection{Atomic SOC of $\sigma$ and $\pi$ bands}
\label{asoc}
We consider next the atomic SOC that mixes the $p$ orbitals, hence, coupling the $\pi$ and $\sigma$ bands: 
\[
\hat H_{\rm{SOC}} = i\lambda^p_{\rm{soc}} \sum_{\alpha=A,B}  \epsilon_{ijk}\hat s_k |p_i^\alpha\rangle\langle p_j^\alpha|+\rm{h.c.}, 
\]
where $\lambda_{\rm{soc}}^p$ is the spin-orbit coupling parameter for the $p$ orbitals and $\hat s_k$ is the $k$-Pauli matrix, 
acting on the spin degree of freedom. 
The action of $\hat H_{\rm{SOC}}$ over the $p_z^\alpha$ orbitals ($\alpha = A,B$) is, 
\[
\hat H_{\rm{SOC}} |p_z^\alpha\rangle\otimes|s_z\rangle = 
\lambda^p_{\rm{soc}}\sqrt{2}s_z|\overline{1 -s_z}\rangle^\alpha, 
\]
where the bar over the eigenstate ($\overline\phi$) indicates opposite spin.
The fist order correction to the $\pi$ orbitals of Eq. (\ref{eq0})  is then:  
\begin{eqnarray}
|\varphi^{(1)}_{\rm AT} \rangle&=&  
\sum_{s,i}\frac{\langle  \phi^s_{i} |\hat H_{\rm{SOC}} |  \varphi^{0} \rangle }{\varepsilon_\sigma^s}  |  \phi^s_{i} \rangle   \nonumber\\
&=&- \sqrt{2}c_A \hat s_z\lambda^p_{\rm{soc}} 
\delta_{s_z,-\tau} \left[ 
\frac{\sin{\gamma}}{\varepsilon^+_\sigma} |\overline \phi^+_{BA} \rangle +
\frac{\cos{\gamma}}{\varepsilon^-_\sigma} |\overline \phi^-_{BA} \rangle\right]
\nonumber \\&&-
 \sqrt{2} c_B \hat s_z\lambda^p_{\rm{soc}}\delta_{s_z,\tau} \left[ \frac{\sin{\gamma}}{\varepsilon^+_\sigma} |\overline \phi^+_{AB} \rangle +
\frac{\cos{\gamma}}{\varepsilon^-_\sigma} |\overline \phi^-_{AB} \rangle\right].
\nonumber 
\end{eqnarray}
Projecting over the terms that yield finite angular momentum correction, we obtain: 
\begin{equation}
\mathcal {\hat P} |\varphi^{(1)}_{\rm AT} \rangle=  
\tau \alpha_{\rm AT}
\left( c_A   |\overline {1-\tau} \rangle^A 
- c_B  |\overline {1\tau} \rangle^B
\right), 
\label{eqs1}
\end{equation} 
with $\mathcal {\hat P} = |11\rangle\langle11| + |1-1\rangle\langle1-1|$ and we have defined the $\sigma$-band 
mixing coefficient, \[
\alpha_{\rm AT} \equiv \sqrt{2} \lambda_{\rm soc}^p\left(\frac{\sin^2{\gamma}}{\varepsilon^+_\sigma}   
+  \frac{\cos^2{\gamma}}{\varepsilon^-_\sigma}\right). 
\]
Formally, the correction given above is similar to that of  (\ref{eqdcorr}), 
accounting for the chirality of the Dirac electrons.
Employing the same arguments as in Sec. \ref{pi} to Eq. (\ref{eqs1}), we conclude that spin-up electrons in the CB
couple to a clockwise rotating $\sigma$ orbital, and the converse for spin-down electrons. CPT symmetry 
ensures that spin-down (-up)  holes couple to (counter)-clockwise orbitals  [see Fig. \ref{figSrot}(b)]. 
\subsubsection{Bychkov-Rashba SOC of $\pi$- and $\sigma$-bands}
\label{app4}
We can express the Bychkov-Rashba Hamiltonian microscopically as the coupling of the $p_z$ and $s$ orbitals, 
\[
\hat H_{\rm{BR}} = \lambda_z \hat {\vec L} \hat{\vec s} \sum_{\alpha=A,B} |s^\alpha\rangle\langle p_z^\alpha| +\rm{H.c.} 
\]
Since $\lambda_z$ is a very small energy scale, we
can safely consider it small with respect to $\varepsilon_\pm$, and treat it perturbatively.
The fist-order correction to the $\pi$ orbitals of Eq. (\ref{eq0})  is then:  
\begin{eqnarray}
|\varphi^{(1)}_{\rm BR}\rangle &=& 
-i \lambda_z\tau c^A \hat {\vec L} \hat{\vec s} \left( \frac{ \cos{\gamma}}{\varepsilon_\sigma^+} | \phi_{AB}^+\rangle
-\frac{\sin{\gamma}}{\varepsilon_\sigma^-} | \phi_{AB}^-\rangle
\right) \nonumber \\&-& i \lambda_z\tau
c^B \hat {\vec L} \hat{\vec s} \left(  \frac{\cos{\gamma}}{\varepsilon_\sigma^+} | \phi_{BA}^+\rangle
-\frac{\sin{\gamma}}{\varepsilon_\sigma^-} | \phi_{BA}^-\rangle \right)
. \nonumber
\end{eqnarray}
Using $\hat {\vec L} \hat{\vec s} = ( \hat L_+\hat s_+ + \hat L_-\hat s_+)/2 + \hat L_z \hat s_z$ and that 
$(\hat L_+\hat s_+ + \hat L_-\hat s_+) |1\pm\tau\rangle^\alpha = 2\delta_{s_z\mp\tau}|p_z^\alpha\rangle$, 
projecting onto the states with finite $m_l$, 
we obtain: 
\begin{equation}
\mathcal{\hat P}|\varphi^{(1)}_{\rm BR}\rangle = 
-i\alpha_{\rm BR}\tau
\left( c_A  | 1\tau\rangle^B \delta_{s_z,-\tau} +c_B | 1 -\tau\rangle^A\delta_{s_z,\tau} \right),
\label{eqEF}
\end{equation}
where we have defined: 
\[
\alpha_{\rm BR}  \equiv \lambda_z \left(\frac{1}{\varepsilon_\sigma^-}-\frac{1}{\varepsilon_\sigma^+}\right)
 \sin{\gamma} \cos{\gamma} = \frac{\sqrt{2}\lambda_z}{3V_{sp\sigma}}
.
\]

\subsubsection{Principal Plane Asymmetry Spin-Orbit Coupling}
We consider, for completeness, a general spin-flipping next-NN  SOC related to the absence of horizontal 
reflection \cite{falko,chengLiu}. 
Although there is no terminological consensus, we adopt the acronym PIA, 
which can be used for pseudospin-inversion asymmetry \cite{kochan2} or more generally, for principal plane mirror asymmetry
\cite{kochan2}. 
This term can be associated to the presence of ripples, defects or adsorbates, and it allows for a momentum-dependent 
coupling of the $p_z^\alpha$ and $p_{x,y}^\beta$-orbitals, hence, allowing coupling between the $\pi$ and $\sigma$ bands. 
The orbital part of the coupling  
takes the general form $ i\tau \beta$, with 
$\beta
= 3n^{B}(V_{pp\sigma}-V_{pp\pi})/4$ real, as detailed in Appendix \ref{appf}: 
\[
\hat H_{\rm PIA} = i\tau \beta \left(|1 \tau\rangle^B\langle p_z^A|\delta_{s_z,\tau} + 
|1 -\tau\rangle^A\langle p_z^B|\delta_{s_z,-\tau} \right)
+ {\rm H.c.}
\]
Near the DPs, we employ perturbation theory and obtain the correction due to PIA SOC:
\begin{eqnarray}
|\varphi^{(1)}_{\rm PIA} \rangle&=&  
\sum_{s,i}\frac{\langle  \phi^s_{i} |\hat H_{\rm{SOC}} |  \varphi^{0} \rangle }{\varepsilon_\sigma^s}  |  \phi^s_{i} \rangle   \nonumber\\
&=&
-i\tau\beta c_A \delta_{s_z,\tau} \left[ \frac{\sin{\gamma}}{\varepsilon_\sigma^+}| \phi_{AB}^+\rangle + 
\frac{\cos{\gamma}}{\varepsilon^-_\sigma} | \phi^-_{AB} \rangle\right]\nonumber \\
&& -i\tau\beta c_B \delta_{s_z,-\tau}\left[ \frac{\sin{\gamma}}{\varepsilon_\sigma^+}| \phi_{BA}^+\rangle + 
\frac{\cos{\gamma}}{\varepsilon^-_\sigma} | \phi^-_{BA} \rangle\right], 
\nonumber 
\end{eqnarray}
which results, projecting over the relevant terms, on a first order correction similar to that of Bychkov-Rashba 
(\ref{eqEF}), only differing in the prefactor, 
\begin{equation}
\hat{\mathcal{P} } |\varphi^{(1)}_{\rm PIA} \rangle = i\tau\alpha_{\rm PIA} 
\left( c_A |1\tau^B\rangle \delta_{s_z,\tau} +c_B |1-\tau^A\rangle \delta_{s_z,-\tau}\right),
\label{eqPIA1}
\end{equation}
with:
\begin{equation}
\alpha_{\rm PIA} 
= \frac{3n^{B}(V_{pp\sigma}-V_{pp\pi})}{4} \left(
\frac{\sin^2{\gamma}}{\varepsilon_\sigma^+} +\frac{\cos^2{\gamma}}{\varepsilon_\sigma^-}
\right).
\label{eqPIA}
\end{equation}

\subsection{Angular momentum contribution of the $\sigma$ band}
We first evaluate the in-plane corrections  $\langle L_{x,y}\rangle_\sigma$, taking into account the first-order 
corrections due to atomic SOC given in  (\ref{eqs1}). 
Using that  $\sqrt{2}\hat L_x|1,\pm\tau\rangle^\alpha = |p_z^\alpha\rangle$ and 
 $\sqrt{2}\hat L_y|1,\pm\tau\rangle^\alpha = \pm i\tau|p_z^\alpha\rangle$, we notice that the only surviving 
first-order correction is along the zigzag direction:
\begin{eqnarray}
\langle \hat L_x \rangle_{\rm AT} 
&=& 2{\rm Re}{ \overline{\langle \varphi_\pm^{(0)} |\hat L_x|\varphi_{\rm AT}^{(1)} \rangle}} =
\pm2\sqrt{2}\alpha_{\rm AT} \tau \sigma_z^0, \nonumber \\ 
\langle \hat L_y \rangle_{\rm AT}  &=&0, 
\end{eqnarray}
where the overbar indicates that we have averaged over spin states. 
This in-plane anisotropy is due to the geometry of the lattice, with a propagating $p_x$ orbital 
mode at the DPs. Recall that $p_x$ orbitals have a finite-$L_x$  contribution. 
The Bychkov-Rashba correction  (\ref{eqEF}) and the PIA correction
yield similar contributions: 
\begin{eqnarray}
\left(
\begin{array}{c} \langle \hat L_x \rangle_{\rm BR}\\ \langle \hat L_y \rangle_{\rm BR}
\end{array}
\right) = \mp 2\sqrt{2} s_z\alpha_{\rm BR} 
\left(
\begin{array}{c}- \sin{\varphi_q} \\ \cos{\varphi_q} 
\end{array}
\right).
\label{eqBRL}
\end{eqnarray}
where we have used $\langle \hat L_\alpha \rangle_{\rm BR} = 2 {\rm Re}{\{ 
\langle \varphi_\pm^{(0)}|\hat L_\alpha |\varphi_{\rm BR}^{(1)} \rangle
\}}$, and likewise, we obtain: 
\begin{eqnarray}
\left(
\begin{array}{c} \langle \hat L_x \rangle_{\rm PIA}\\ \langle \hat L_y \rangle_{\rm PIA}
\end{array}
\right) = \mp 2\sqrt{2}\tau s_z\alpha_{\rm PIA} 
\left(
\begin{array}{c}- \sin{\varphi_q} \\ \cos{\varphi_q} 
\end{array}
\right). 
\label{eqPIAL}
\end{eqnarray}
We note that   the Rashba and PIA terms   break  parity symmetry 
and, as a consequence, the corrections in Eqs. (\ref{eqBRL}) and (\ref{eqPIAL}) 
resemble  the {\it pseudomomentum} and {\it pseudotorque} defined in (\ref{eqPLpar}) and (\ref{eqLpar}), respectively. 

We now consider the axial correction due to $\sigma$-band mixing, $\langle \hat L_z \rangle_\sigma$. 
Equations ({\ref{eqs1}}), ({\ref{eqEF}}), and (\ref{eqPIA1}) yield three different second-order contributions, 
which result into three terms proportional to $\tau \langle\sigma_z\rangle$, 
owing to the SL symmetry breaking, 
\[
\langle\hat L_z\rangle_\sigma \simeq \left(
(\alpha_{\rm BR} + \alpha_{\rm PIA})^2\ -\alpha_{\rm AT}^2 \right)\tau\sigma_z^0, 
\]
where the negative sign in front of the intrinsic SOC  term  indicates that spin-up (-down) 
electrons couple to (anti)-clockwise rotating $p$ orbitals within the $\sigma$ band, 
and the converse for the holes [see Fig. \ref{figSrot}(b)].



For completeness, we evaluate higher-order corrections along the  armchair direction, $y$. 
Involved calculations yield third-order corrections,  (see Appendix \ref{appd}):  
\begin{equation}
\langle \hat L_y \rangle_\sigma \simeq \alpha_{\rm AT}^2 \alpha_{\rm BR}
\left( 
\langle\hat{\vec\sigma}_\tau\hat{\vec s}\rangle   
+\tau \sigma_z^0 \langle \hat s_y\rangle  
\right) + 
\alpha_{\rm AT}^3\tau \langle \hat s_y\rangle. 
\label{eqLy}
\end{equation} 
In presence of inversion symmetry,  $\alpha_{\rm BR} = \alpha_{\rm PIA}=0$, 
the corrections along the armchair direction would only appear to third order, 
reflecting the intrinsic peculiarities of the lattice structure. 

We have thus encountered striking anisotropies in the angular momentum corrections, which are first order 
along $\hat x$, second order along $\hat z$, and third order or higher along the armchair direction.  
Taking into account the correction of the $\pi$ bands given in (\ref{eqLzPi}), we can summarize our result as: 
\begin{eqnarray}
\langle \hat L_x\rangle &\simeq& \pm 2\sqrt{2}[\alpha_{\rm AT}\tau\sigma_z^0 +
(\alpha_{\rm PIA}+\alpha_{\rm BR})s_z\tau\sin{\varphi_q}],
\nonumber \\
\langle \hat L_y\rangle &\simeq& \mp 2\sqrt{2} (\alpha_{\rm PIA}+ \alpha_{\rm BR})s_z\tau\cos{\varphi_q}, 
\nonumber \\
\langle \hat L_z\rangle &\simeq& \left[ \frac{\lambda_I}{\lambda_{\rm soc}^d} 
- \alpha_{\rm AT}^2
+ (\alpha_{\rm BR}  + \alpha_{\rm PIA})^2 
\right]\tau\sigma_z^0 . 
\label{central}
\end{eqnarray}
Although in an ideal, single-particle picture  $\langle\tau\sigma_z^0\rangle =0$ and $\langle\tau s_z^0\rangle =0$
symmetry breaking terms would allow to resolve inherent internal structure of the Dirac carriers. 
For instance, electron-spin resonance on a graphene-doped sample would allow to address directly the angular momentum correction. 
Since the value of the gap is known, $2\lambda_I \simeq 42$ $\mu$eV\cite{jonas,BanszerusPRL20}, 
resolving  the $g$ tensor along the main crystallographic directions  could be employed to find the value of atomic SOC, $\lambda_{\rm soc}^{p,d}$ 
\cite{exp}
or the magnitude of the Stark parameter, $\lambda_z$.  

\section{Conclusions }
\label{con}
We have considered a Dirac Hamiltonian in graphene with a mass term and 
we have obtained the finite angular momentum corrections along the main crystallographic axis.
Intrinsic  SOC causes the positive-energy carriers (electrons) 
in one SL with spin `up' to couple to anti-clockwise rotating orbitals, 
whereas those with spin `down' couple to clockwise rotating orbitals, 
with the converse occurring for negative-energy quasiparticles (holes).  
This
confirms the CPT symmetry in the Hamiltonian, and, ultimately, reflects the chirality of the Dirac electrons.  
We have developed an intuitive connection between angular momentum and SL spin, where the 
axial quantization is given in terms of the sum of usual $L_z$ operator and $\nu\sigma_z$, with $\nu$ being the winding number. 
This connection is very important when establishing the selection rules that dictate how Dirac carriers couple to other 
carrying angular momentum quasiparticles, such as photons. 
Corrections to the in-plane momentum  are related to a pseudovector whose torque is perpendicular to the in-plane SL spin. 
In the absence of external SOC,  we find first-order corrections along the zigzag direction, 
owing to the propagating $p_x$ orbitals at DPs, 
whereas no corrections are observed in the armchair direction (up to third order).
Whereas the intrinsic correction is associated with the pseudoscalar $\tau \sigma_z$, the 
extrinsic correction is related to an in-plane pseudovector perpendicular to in-plane SL spin  $\sigma_\parallel^\tau$. 
The angular momentum correction anisotropy presented here has been confirmed
in recent angle-resolved electron-spin resonance experiments \cite{exp}. 

{\it Acknowledgements: }We acknowledge support by the Bundesminsterium f\"ur Forschung und Technologie (BMBF) through the
`Forschungslabor Mikroelectronik Deutschland (ForLab)'. We thank L. Tiemann, R. H. Blick and 
T. Schmirander for fruitful discussions.

\appendix
\section{$\hat{\vec\sigma}_\tau^\parallel$ is not a pseudovector}
\label{appa}
A pseudovector or axial vector transforms like a polar vector under rotations, but gains a sign flip under improper rotations, 
such as parity inversion.
Here, we show that $ \hat{\vec \sigma}_\tau\equiv (\tau \hat \sigma_x,  \hat\sigma_y, \hat\sigma_z)$  
transforms as a polar vector, in spite of being expressed in terms of the  Pauli matrices.

Under inversion symmetry, $\mathcal{P}$, the sublattices $A$ and $B$ are switched, as well as the valleys, and hence, 
the Pauli matrices transform as 
$\mathcal{P}: (\hat\sigma_x,\hat\sigma_y,\hat\sigma_z) \to  (\hat\sigma_x,-\hat\sigma_y,-\hat\sigma_z). $
Noting that the valleys are also switched, $\mathcal{P} \tau = -\tau$, we obtain:
\[
\mathcal{\hat P} \hat{\vec\sigma}_\tau = (-\tau\hat\sigma_x,-\hat\sigma_y,-\hat\sigma_z).
\]
We conclude that $\hat{\vec\sigma}_\tau$ is a polar vector, as well as the in-plane component,
$\hat{\vec\sigma}_\tau^\parallel$.  

Hence, the first term of Eq. (\ref{eqH0}) is a scalar: it is straight forward to 
see that it is invariant under parity, as both $\vec q$ and $\vec\sigma_\tau$ change sign under inversion transformation.
The Semenov term, $\varepsilon_{AB}\sigma_z$ is, however, a pseudoscalar, as opposed to the parity-preserving 
Kane-Mele term, $\lambda_I\tau \hat s_z\hat \sigma_z$.


It is tempting to express the Hamiltonian of Eq. (\ref{eqH0}) as a scalar product of
$ \hat{\vec\sigma}_\tau  = (\tau \sigma_x, \sigma_y ,\sigma_z)$
and $ \hat{\vec q}_{\rm 3d}=(q_x,q_y, \hat \lambda /\hbar v_F )$ by identifying $\hat q_z = \hat \lambda/\hbar v_F$. 
However, $\hat{\vec q}_{\rm 3d}$ is  neither a  polar nor an axial vector, and the third 
component can not be associated with a momentum, leading to unphysical interpretations.

\section{Constant of motion in Dirac Hamiltonians}
\label{app1b}
In the search for a  generalization of Eq. (\ref{eqLz}) in other Dirac Hamiltonians, we focus on bilayer graphene. 
As MacCann {\it et al.} demonstrated, \cite{mccannBLG} quasiparticles in bilayer graphene can be described by using 
the effective Hamiltonian, 
\[
\hat H_{\rm BG} = -\tau\frac{\hbar^2 q^2}{2m^*}\vec\sigma_\tau^\parallel \vec n, \quad \vec n = (\cos{2\varphi_q,\sin{2\varphi_q}})
\]
where the resulting eigenstates gain a phase shift of $2\pi$ under an adiabatic rotation of a closed contour in momentum space, 
corresponding to a Berry phase of 2$\pi$ \cite{Novoselov2006} (or a winding number $\nu = 2$).
The time evolution of $\hat \sigma_z$ reads as: 
\begin{eqnarray}
\frac{{\rm d} \hat \sigma_z}{{\rm d} t} &=& 
\frac{-i\tau\hbar q^2}{ 2m^*} \left\{ \left[ \tau\hat \sigma_x \cos{2\varphi_q} + \hat\sigma_y \sin{2\varphi_q} , \hat \sigma_z\right]
\right\} \nonumber \\ &=& \frac{\tau\hbar q^2}{m^*} \left(\tau\hat\sigma_y\cos{2\varphi_q} - \hat\sigma_x\sin{2\varphi_q} \right) , 
\end{eqnarray}
and likewise, we have for $\hat L_z$: 
\begin{eqnarray}
\frac{{\rm d} \hat L_z}{{\rm d} t} &=& 
\frac{-i\tau\hbar^2}{ 2m^*} 
\left\{ \left[ \tau\hat \sigma_x (\hat q_x^2 -\hat q_y^2) + 2\hat\sigma_y \hat q_x\hat q_y , \hat x\hat q_y -\hat y\hat q_z\right]
\right\}  \nonumber \\ &=& \frac{\tau\hbar^2 q^2}{m^*} \left(\tau \hat\sigma_x\sin{2\varphi_q} - \hat\sigma_y\cos{2\varphi_q} \right)= 
-\hbar\tau\dot{\sigma}_z,
\end{eqnarray}
which is, combining (B1) and (B2), 
\[
\frac{{\rm d}}{{\rm d} t}( L_z + \hbar \tau\sigma_z ) = 0, 
\]
resulting in (\ref{eqLz}) for $\nu =2$. 
\section{ The hybridized $\pi$-bands } 
\label{appb}
\begin{figure}[!hbt]
\parbox[c]{.99\linewidth}
{
\includegraphics[angle=0, width=.50\linewidth]{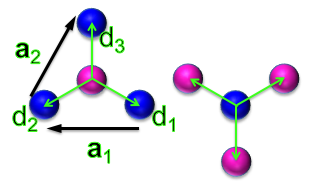}}\\
\caption{ $A$ (blue) and $B$ (magenta) sublattices, with the NNs directive cosines, 
$\vec d_{1,2}^{BA} =-\vec d_{1,2}^{AB} =(\pm\sqrt{3}/2, -1/2 )$  and 
$\vec d_3^{AB} - \vec d_3^{BA} = (0,1)$ and the primitive vectors, 
$\vec a_{1,2}$. 
}
\label{figAB}
\end{figure}

We consider the unperturbed eigenstates in terms of $p_z$ orbitals (\ref{eq0}).  
Using perturbation theory within the two-center Slater-Koster approximation \cite{slater}
we obtain the non-zero $d$-orbital contribution, 
\begin{eqnarray}
|\varphi^{(1)}_d\rangle &\simeq& 
-\frac{c_A}{\varepsilon_{pd}}\left( 
V_{xz,z}^{BA}(k) 
|d_{xz}^B\rangle + 
V_{yz,z}^{BA}(k)
|d_{yz}^B\rangle\right)
\nonumber \\ &&
-\frac{c_B}{\varepsilon_{pd}}\left(  
V_{xz,z}^{AB}(k) 
|d_{xz}^A\rangle + 
V_{yz,z}^{AB}(k) 
|d_{yz}^A\rangle \right). 
\nonumber
\label{eqAd1}
\end{eqnarray}
The matrix elements  $V_{iz,z}^{\alpha \beta}(k) \equiv \sum_i\langle d_{xz}^\alpha|\hat V|p_z^\beta \rangle \exp{(i\vec k\vec d_i)}$  
are to be evaluated near the DPs,  {\it i.e.}, at $\vec k = \vec q + \vec K^{(\prime)}$, with $\vec K = -\vec K^\prime = 
(4\pi/3\sqrt{3}a,0)$, and $a$ being the distance between two NN $^12$C atoms: 
\[  (k_x, k_y) =  \left( \frac{4\pi\tau}{3\sqrt{3} a} +q_x, q_y\right). 
\]
The $p$-$d$ hopping is described in terms of the NN directive
cosines, $\vec d_i = (l_i,m_i)$ as 
depicted in Fig. \ref{figAB}. Those are $\vec d_{1,2} = (\pm\sqrt{3}/2, -1/2 )$  and $\vec d_3 = (0,1)$,  
 giving, 
{\it e.g.},  for the coupling of $d_{yz}^B$ and NN $p_z^A$ orbitals:
\begin{eqnarray}
V_{xz,z}^{BA}(k)  &=&V_{pd\pi}
\sum_i  
l_i e^{i\vec k \vec d_i} 
\nonumber \\ &=& 
\frac{\sqrt{3}}{2} V_{pd\pi} \left( e^{ik_x m_1} - e^{i k_x  m_2} \right)e^{-i\frac{q_ya}{2}} 
\nonumber \\ &=& 
i\sqrt{3}V_{pd\pi} e^{-i\frac{q_ya}{2}} \sin{\left(\frac{2\pi\tau}{3} +q_x a\frac{\sqrt{3}}{2}\right) }
\nonumber \\&\simeq&
 i\tau \frac{{3V_{pd\pi}}}{2} \left( 1 -  \frac{ aq}{2} e^{i\tau\varphi_q} \right),  
\label{eqVdxzpz}
\end{eqnarray}
where we have used that $V_{xz,z} = lV_{pd\pi}$. 
Likewise, we have:  
\begin{eqnarray}
V_{yz,z}^{BA}(k)  &=&V_{pd\pi}
\sum_i m_i  e^{i\vec k \vec d_i}    \nonumber \\ &=&  V_{pd\pi} \left(  
\frac{-1}{2}  \left( e^{i k_x \sqrt{3}/2} + e^{-i k_x\sqrt{3}/2 } \right)e^{-i\frac{q_y}{2}} +   e^{iq_y}
\right)
\nonumber \\&\simeq&
\frac{{3V_{pd\pi} }}{2}\left( 1 + \frac{ aq}{2} e^{i\tau\varphi_q} \right).  
\label{eqVdyzpz}
\end{eqnarray}
 It is worth noting that under sublattice swap, the matrix elements change sign, that is, 
$ V_{iz,z}^{\alpha\beta} (k) = \sum_i  V_{iz,z}^{\alpha\beta}  e^{i(k_x l_i + k_ym_i)} = 
-\sum_i V_{iz,z}^{\beta\alpha}  e^{-i(k_x l_i+ k_ym_i)} $, since the directive cosines change sign 
(see Fig. \ref{figAB}): 
\[
V_{iz,z}^{AB} (k)  =  -[V_{iz,z}^{BA} (k)]^* = -V_{z,iz}^{AB} (k). 
\]
This has important consequences, implying that the topological character of the $\pi$ bands is preserved.
To lowest order, that is, neglecting the linear in momentum terms, we obtain: 
\begin{eqnarray}
|\varphi^{(1)}_d\rangle &\simeq& -\frac{3V_{pd\pi}}{2\varepsilon_{pd}}
\left[c_A\left( 
|d_{yz}^B\rangle  +i \tau|d_{xz}^B\rangle
\right)   
\right.\nonumber \\ & &\left.
+c_B
\left( 
-|d_{yz}^A\rangle  +i \tau|d_{xz}^A\rangle
\right) 
\right]    
\nonumber \\ & =&-
\frac{3i\tau V_{pd\pi}}{\sqrt{2}\varepsilon_{pd}}
\left(c_A  |2 \tau\rangle^B + c_B|2 -\tau\rangle^A\right), 
\nonumber \\ 
\end{eqnarray}
where in the last step we have used the angular momentum representation, $|l,m_l\rangle$, 
with $l=2$, and  
\[
|2\pm 1\rangle = \frac{1}{\sqrt{2}} \left( |d_{xz}\rangle\pm i|d_{yz}\rangle \right). 
\]
\section{The hybridized $\sigma$ bands near DPs}
\label{appc}
We consider the $sp^2$ hybridization via $\sigma$ bonds near the DPs.  
The low-energy Hamiltonian can be evaluated within a reduced Hilbert space spanned by $\{s^\alpha,p_{x}^\beta,p_y^\beta\}$, 
$\alpha\ne \beta$, as dominant hopping within 
the $\sigma$ band occurs between NN $s$- and $p_{x,y}$-orbitals \cite{min2006intrinsic}.
The  hoppings between an $s$-orbital in SL $A$ and $p_{x}$ ($p_y$) in the  SL $B$ 
is given by $l_i V_{sp\sigma}$ ($m_i V_{sp\sigma}$). 
Hence, we can employ the results of the previous section, where only the prefactor changes. 
Using Eq. (\ref{eqVdxzpz}), we obtain: 
\begin{eqnarray}
V_{s,p_x}^{BA}(k) &   
\simeq & V_{sp\sigma} \sum_i l_i e^{i\vec k \vec d_i} = 
i\tau\frac{{3 V_{sp\sigma}}}{2} \left( 1 -\frac{aq}{2}e^{i\tau\varphi_q}\right)\nonumber 
\end{eqnarray}
Likewise,  we have:
\begin{eqnarray}
V_{s,p_y}^{BA}(k) 
&\simeq& \frac{3}{2}V_{sp\sigma} \left( 1 +\frac{a}{2} ( \tau q_x + i q_y)\right), 
\nonumber
\end{eqnarray}
and it follows, using symmetry arguments, $V_{s,p_x}^{AB}(k) = -[V_{s,p_x}^{BA}(k)]*$, that is:
\[
V_{s,p_x}^{AB}(k)
\simeq  \frac{3i\tau}{2}V_{sp\sigma} \left( 1 -\frac{a}{2} ( \tau q_x - i q_y)\right), \]\[
V_{s,p_y}^{AB}(k)
\simeq  -\frac{3}{2}V_{sp\sigma} \left( 1 +\frac{a}{2} ( \tau q_x - i q_y)\right).
\]
In the basis spanned by $\{ s^A, |1,\tau\rangle^B\}, |1,-\tau\rangle^B\}$  with $|1,\pm\tau\rangle^B\equiv 
(|p_x\rangle^B \mp i\tau |p_y\rangle^B)/\sqrt{2}$, we have: 
\begin{equation}
\hat H_h^{BA} = \left(
\begin{array}{ccc} \varepsilon_s & 3i\tau V_{sp\sigma}/\sqrt{2} &0\\ -3i\tau V_{sp\sigma}/\sqrt{2}&\varepsilon_p &0\\0&0&\varepsilon_p
\end{array}
\right), {\rm { }}
\label{eqHBAs}
\end{equation}
with eigenvalues: 
\[
 \varepsilon_\sigma^\pm =\frac{\varepsilon_s+\varepsilon_p}{2}\pm \sqrt{\frac{(\varepsilon_{s}-\varepsilon_p)^2}{4} +\frac{9V_{sp\sigma}^2}{2}},
 \  \varepsilon_0 = \varepsilon_p, 
\]
and corresponding eigenvectors, with $\phi^0_{AB} = |1-\tau\rangle^B$ and 
\begin{equation}
\phi^+_{AB}= \left( \begin{array}{c} i\tau\cos{\gamma} \\ \sin{\gamma} \\0 
\end{array} \right), \quad
 \phi^-_{AB}= \left( \begin{array}{c} -i\tau\sin{\gamma} \\ \cos{\gamma}\\0 
\end{array} \right),  
\label{EqBAev}
\end{equation}
where we have defined
\[
\gamma = \arctan{  \left[
\frac{3V_{sp\sigma}}{\sqrt{2} (\varepsilon_\sigma^+-\varepsilon_p)}  \right]}. 
\]

Likewise, we have that in the basis spanned by $\{ |s\rangle^B, |1,-\tau\rangle^A,|1,\tau\rangle^A\}$, 
the hopping Hamiltonian is formally identical to that of Eq. (\ref{eqHBAs}), 
yielding  same eigenvalues, and similar eigenvectors:  
\begin{equation}
\phi^+_{BA}= \left( \begin{array}{c} i\tau\cos{\gamma} \\ \sin{\gamma} \\0 
\end{array} \right), \quad
 \phi^-_{BA}= \left( \begin{array}{c} -i\tau\sin{\gamma} \\ \cos{\gamma} \\0
\end{array} \right), 
\label{eqABev}
\end{equation}
with  $\phi^0_{AB} = |1\tau\rangle^A$. 
\section{PIA coupling}
\label{appf}
We consider now the  coupling induced by horizontal plane mirror asymmetry, PIA. 
The term arises in the presence of adsorbates, \cite{falko} ripples \cite{kochan,kochan2} 
or other defects, and causes a coupling between the $p_z$ and the 
NNs $p_{x,y}$ orbitals \cite{chengLiu}. 
We account for this term assuming a vertical relative component of one lattice with respect to the other,  
with a directive cosinus given by $n_i^{AB} = n^B$ $\forall$ $i$. 
Employing the two-center Slater-Koster approximation \cite{slater}, the relevant orbital matrix elements near DPs are 
$V^{BA}_{\eta,z}(q) \equiv  \langle p_\eta^B|\hat V|p_z ^A \rangle $, with: 
\begin{eqnarray}
V^{BA}_{x,z}(q) &=&  (V_{pp\sigma}-V_{pp\pi})\sum_i n_il_i e^{i\vec k \vec d_i }= \nonumber \\ &=&
  i\tau \frac{3n^{B}}{2}(V_{pp\sigma}-V_{pp\pi}) \left( 1 -  \frac{ aq}{2} e^{i\tau\varphi_q} \right), \nonumber\\ 
V^{BA}_{y,z}(q) &=&  (V_{pp\sigma}-V_{pp\pi})\sum_i n_i m_i e^{i\vec k \vec d_i }= \nonumber \\ &=&
  \frac{3n^{B}}{2}(V_{pp\sigma}-V_{pp\pi}) \left( 1 +  \frac{ aq}{2} e^{i\tau\varphi_q} \right),\nonumber
\end{eqnarray}
where we have employed the procedure of Eqs. (\ref{eqVdxzpz}) and (\ref{eqVdyzpz}), and $n^{B}$ is a sublattice $B$
vertical displacement. $V^{AB}_{\eta,z}(q)$ are trivially obtained, using $V^{AB}_{\eta,z}(q,n^{B}) = -(V^{BA}_{\eta,z}(q,n^{A}))^*$. 
The allowed SOC term is then of the form $(\hat L_+\hat s_-+\hat L_- \hat s_+)/2$, which results into
the coupling of the $p_z$ orbitals to the $\sigma$ band is given by: 
\[
\frac{1}{2}\langle 1\tau |^B \hat V|p_z ^A \rangle\otimes (\hat s_x +i\tau \hat s_y) = i\tau \delta_{s_z,\tau}
\frac{3n^{B}}{4}(V_{pp\sigma}-V_{pp\pi}), 
\]
\[
\frac{1}{2}\langle 1-\tau |^A \hat V|p_z ^B \rangle \otimes (\hat s_x-i\tau \hat s_y) =  
i\tau \delta_{s_z,-\tau} \frac{3n^{A}}{4}(V_{pp\sigma}-V_{pp\pi}),  
\]
where the spin degree of freedom has been explicitly evaluated in terms of the valley, to ensure PT symmetry. 

\section{Third order corrections}
\label{appd}
We consider the  $\hat H_{\rm BR} + \hat H_{\rm soc}$ and calculate the corrections over $|\varphi^{(1)}_{\rm soc}\rangle$ and 
$|\varphi^{(1)}_{\rm BR}\rangle$, which are the second order corrections to the WF: 
\begin{eqnarray}
|\varphi^{(2)} \rangle 
&\simeq& \alpha_{\rm soc}^2
\left[
c_A \langle \tau \hat s_x + i\hat s_y\rangle^A |p_z^A\rangle -c_B \langle -\tau \hat s_x + i\hat s_y\rangle^B |p_z^B\rangle
\right]\nonumber \\  &-&
i\alpha_{\rm BR} \alpha_{\rm soc}\left[
c_A \langle -\tau \hat s_x + i\hat s_y\rangle^B |p_z^B\rangle + \right. \nonumber \\ &&\left. 
 c_B\langle \tau \hat s_x + i\hat s_y\rangle^A |p_z^A\rangle
\right], 
\nonumber 
\end{eqnarray}
where $\langle\dots \rangle^\alpha$ indicates that the expectation value is to be calculated in SL $\alpha$. 
The next order correction to angular momentum along $y$ would be 
$ 2 {\rm Re}{\left\{\langle \varphi^{(1)}_\sigma |\hat L_y|\varphi^{(2)}\rangle\right\}}$, yielding: 
\begin{eqnarray}
\langle \hat L_y\rangle &\simeq& 
\alpha_{\rm soc}^3\tau (\langle \hat s_y\rangle^A |c_A|^2 + \langle \hat s_y\rangle^B  |c_B|^2) + 
\nonumber \\ &&
\alpha_{\rm soc}^2\alpha_{\rm BR} \left(\tau c_A^*c_B \langle \hat s_x\rangle^A 
+ \tau c_Ac_B^* \langle \hat s_x\rangle^B  
\right.\nonumber \\ &&\left.
+ i  c_A^*c_B \langle \hat s_y\rangle^A -i  c_Ac_B^* \langle \hat s_y\rangle^B \right). 
\end{eqnarray}
In absence of spin-SL coupling, we can consider $\langle \hat s_i\rangle^A = \langle \hat s_i\rangle^B = \langle \hat s_i\rangle$, $i=x,y$. 
Using that $|c_A|^2  + |c_B|^2 \simeq 1$ and (\ref{eq0}),  we obtain (\ref{eqLy}).

\bibliography{bibliography}

\providecommand{\noopsort}[1]{}\providecommand{\singleletter}[1]{#1}%
\begin{thebibliography}{40}%
\makeatletter
\providecommand \@ifxundefined [1]{%
 \@ifx{#1\undefined}
}%
\providecommand \@ifnum [1]{%
 \ifnum #1\expandafter \@firstoftwo
 \else \expandafter \@secondoftwo
 \fi
}%
\providecommand \@ifx [1]{%
 \ifx #1\expandafter \@firstoftwo
 \else \expandafter \@secondoftwo
 \fi
}%
\providecommand \natexlab [1]{#1}%
\providecommand \enquote  [1]{``#1''}%
\providecommand \bibnamefont  [1]{#1}%
\providecommand \bibfnamefont [1]{#1}%
\providecommand \citenamefont [1]{#1}%
\providecommand \href@noop [0]{\@secondoftwo}%
\providecommand \href [0]{\begingroup \@sanitize@url \@href}%
\providecommand \@href[1]{\@@startlink{#1}\@@href}%
\providecommand \@@href[1]{\endgroup#1\@@endlink}%
\providecommand \@sanitize@url [0]{\catcode `\\12\catcode `\$12\catcode
  `\&12\catcode `\#12\catcode `\^12\catcode `\_12\catcode `\%12\relax}%
\providecommand \@@startlink[1]{}%
\providecommand \@@endlink[0]{}%
\providecommand \url  [0]{\begingroup\@sanitize@url \@url }%
\providecommand \@url [1]{\endgroup\@href {#1}{\urlprefix }}%
\providecommand \urlprefix  [0]{URL }%
\providecommand \Eprint [0]{\href }%
\providecommand \doibase [0]{http://dx.doi.org/}%
\providecommand \selectlanguage [0]{\@gobble}%
\providecommand \bibinfo  [0]{\@secondoftwo}%
\providecommand \bibfield  [0]{\@secondoftwo}%
\providecommand \translation [1]{[#1]}%
\providecommand \BibitemOpen [0]{}%
\providecommand \bibitemStop [0]{}%
\providecommand \bibitemNoStop [0]{.\EOS\space}%
\providecommand \EOS [0]{\spacefactor3000\relax}%
\providecommand \BibitemShut  [1]{\csname bibitem#1\endcsname}%
\let\auto@bib@innerbib\@empty
\bibitem [{\citenamefont {Novoselov}\ \emph {et~al.}(2005)\citenamefont
  {Novoselov}, \citenamefont {Geim}, \citenamefont {Morozov}, \citenamefont
  {Jiang}, \citenamefont {Katsnelson}, \citenamefont {Grigorieva},
  \citenamefont {Dubonos},\ and\ \citenamefont {Firsov}}]{novoselov2005two}%
  \BibitemOpen
  \bibfield  {author} {\bibinfo {author} {\bibfnamefont {K.~S.}\ \bibnamefont
  {Novoselov}}, \bibinfo {author} {\bibfnamefont {A.~K.}\ \bibnamefont {Geim}},
  \bibinfo {author} {\bibfnamefont {S.~V.}\ \bibnamefont {Morozov}}, \bibinfo
  {author} {\bibfnamefont {D.}~\bibnamefont {Jiang}}, \bibinfo {author}
  {\bibfnamefont {M.~I.}\ \bibnamefont {Katsnelson}}, \bibinfo {author}
  {\bibfnamefont {I.~V.}\ \bibnamefont {Grigorieva}}, \bibinfo {author}
  {\bibfnamefont {S.~V.}\ \bibnamefont {Dubonos}}, \ and\ \bibinfo {author}
  {\bibfnamefont {A.~A.}\ \bibnamefont {Firsov}},\ }\href@noop {} {\bibfield
  {journal} {\bibinfo  {journal} {Nature}\ }\textbf {\bibinfo {volume} {438}},\
  \bibinfo {pages} {197} (\bibinfo {year} {2005})}\BibitemShut {NoStop}%
\bibitem [{\citenamefont {Katsnelson}\ \emph {et~al.}(2006)\citenamefont
  {Katsnelson}, \citenamefont {Novoselov},\ and\ \citenamefont
  {Geim}}]{katsnelson2006chiral}%
  \BibitemOpen
  \bibfield  {author} {\bibinfo {author} {\bibfnamefont {M.~I.}\ \bibnamefont
  {Katsnelson}}, \bibinfo {author} {\bibfnamefont {K.}~\bibnamefont
  {Novoselov}}, \ and\ \bibinfo {author} {\bibfnamefont {A.~K.}\ \bibnamefont
  {Geim}},\ }\href@noop {} {\bibfield  {journal} {\bibinfo  {journal} {Nature
  Physics}\ }\textbf {\bibinfo {volume} {2}},\ \bibinfo {pages} {620} (\bibinfo
  {year} {2006})}\BibitemShut {NoStop}%
\bibitem [{\citenamefont {Geim}\ and\ \citenamefont
  {Novoselov}(2010)}]{geim2010rise}%
  \BibitemOpen
  \bibfield  {author} {\bibinfo {author} {\bibfnamefont {A.~K.}\ \bibnamefont
  {Geim}}\ and\ \bibinfo {author} {\bibfnamefont {K.~S.}\ \bibnamefont
  {Novoselov}},\ }in\ \href@noop {} {\emph {\bibinfo {booktitle} {Nanoscience
  and Technology: A Collection of Reviews from Nature Journals}}}\ (\bibinfo
  {publisher} {World Scientific},\ \bibinfo {year} {2010})\ pp.\ \bibinfo
  {pages} {11--19}\BibitemShut {NoStop}%
\bibitem [{\citenamefont {Nair}\ \emph {et~al.}(2008)\citenamefont {Nair},
  \citenamefont {Blake}, \citenamefont {Grigorenko}, \citenamefont {Novoselov},
  \citenamefont {Booth}, \citenamefont {Stauber}, \citenamefont {Peres},\ and\
  \citenamefont {Geim}}]{Nair08}%
  \BibitemOpen
  \bibfield  {author} {\bibinfo {author} {\bibfnamefont {R.~R.}\ \bibnamefont
  {Nair}}, \bibinfo {author} {\bibfnamefont {P.}~\bibnamefont {Blake}},
  \bibinfo {author} {\bibfnamefont {A.~N.}\ \bibnamefont {Grigorenko}},
  \bibinfo {author} {\bibfnamefont {K.~S.}\ \bibnamefont {Novoselov}}, \bibinfo
  {author} {\bibfnamefont {T.~J.}\ \bibnamefont {Booth}}, \bibinfo {author}
  {\bibfnamefont {T.}~\bibnamefont {Stauber}}, \bibinfo {author} {\bibfnamefont
  {N.~M.~R.}\ \bibnamefont {Peres}}, \ and\ \bibinfo {author} {\bibfnamefont
  {A.~K.}\ \bibnamefont {Geim}},\ }\href {\doibase 10.1126/science.1156965}
  {\bibfield  {journal} {\bibinfo  {journal} {Science}\ }\textbf {\bibinfo
  {volume} {320}},\ \bibinfo {pages} {1308} (\bibinfo {year}
  {2008})}\BibitemShut {NoStop}%
\bibitem [{\citenamefont {Katsnelson}\ and\ \citenamefont
  {Novoselov}(2007)}]{katsnelson2007}%
  \BibitemOpen
  \bibfield  {author} {\bibinfo {author} {\bibfnamefont {M.}~\bibnamefont
  {Katsnelson}}\ and\ \bibinfo {author} {\bibfnamefont {K.}~\bibnamefont
  {Novoselov}},\ }\href {\doibase https://doi.org/10.1016/j.ssc.2007.02.043}
  {\bibfield  {journal} {\bibinfo  {journal} {Solid State Communications}\
  }\textbf {\bibinfo {volume} {143}},\ \bibinfo {pages} {3 } (\bibinfo {year}
  {2007})},\ \bibinfo {note} {exploring graphene}\BibitemShut {NoStop}%
\bibitem [{\citenamefont {Giuliani}\ \emph {et~al.}(2012)\citenamefont
  {Giuliani}, \citenamefont {Mastropietro},\ and\ \citenamefont
  {Porta}}]{GIULIANI2012461}%
  \BibitemOpen
  \bibfield  {author} {\bibinfo {author} {\bibfnamefont {A.}~\bibnamefont
  {Giuliani}}, \bibinfo {author} {\bibfnamefont {V.}~\bibnamefont
  {Mastropietro}}, \ and\ \bibinfo {author} {\bibfnamefont {M.}~\bibnamefont
  {Porta}},\ }\href {\doibase https://doi.org/10.1016/j.aop.2011.10.007}
  {\bibfield  {journal} {\bibinfo  {journal} {Annals of Physics}\ }\textbf
  {\bibinfo {volume} {327}},\ \bibinfo {pages} {461 } (\bibinfo {year}
  {2012})}\BibitemShut {NoStop}%
\bibitem [{\citenamefont {Marino}\ \emph {et~al.}(2015)\citenamefont {Marino},
  \citenamefont {Nascimento}, \citenamefont {Alves},\ and\ \citenamefont
  {Smith}}]{PhysRevX.5.011040}%
  \BibitemOpen
  \bibfield  {author} {\bibinfo {author} {\bibfnamefont {E.~C.}\ \bibnamefont
  {Marino}}, \bibinfo {author} {\bibfnamefont {L.~O.}\ \bibnamefont
  {Nascimento}}, \bibinfo {author} {\bibfnamefont {V.~S.}\ \bibnamefont
  {Alves}}, \ and\ \bibinfo {author} {\bibfnamefont {C.~M.}\ \bibnamefont
  {Smith}},\ }\href {\doibase 10.1103/PhysRevX.5.011040} {\bibfield  {journal}
  {\bibinfo  {journal} {Phys. Rev. X}\ }\textbf {\bibinfo {volume} {5}},\
  \bibinfo {pages} {011040} (\bibinfo {year} {2015})}\BibitemShut {NoStop}%
\bibitem [{\citenamefont {Golub}\ \emph {et~al.}(2020)\citenamefont {Golub},
  \citenamefont {Egger}, \citenamefont {M\"uller},\ and\ \citenamefont
  {Villalba-Ch\'avez}}]{golub}%
  \BibitemOpen
  \bibfield  {author} {\bibinfo {author} {\bibfnamefont {A.}~\bibnamefont
  {Golub}}, \bibinfo {author} {\bibfnamefont {R.}~\bibnamefont {Egger}},
  \bibinfo {author} {\bibfnamefont {C.}~\bibnamefont {M\"uller}}, \ and\
  \bibinfo {author} {\bibfnamefont {S.}~\bibnamefont {Villalba-Ch\'avez}},\
  }\href {\doibase 10.1103/PhysRevLett.124.110403} {\bibfield  {journal}
  {\bibinfo  {journal} {Phys. Rev. Lett.}\ }\textbf {\bibinfo {volume} {124}},\
  \bibinfo {pages} {110403} (\bibinfo {year} {2020})}\BibitemShut {NoStop}%
\bibitem [{\citenamefont {Kane}\ and\ \citenamefont
  {Mele}(2005{\natexlab{a}})}]{kane2005quantum}%
  \BibitemOpen
  \bibfield  {author} {\bibinfo {author} {\bibfnamefont {C.~L.}\ \bibnamefont
  {Kane}}\ and\ \bibinfo {author} {\bibfnamefont {E.~J.}\ \bibnamefont
  {Mele}},\ }\href@noop {} {\bibfield  {journal} {\bibinfo  {journal} {Physical
  Review Letters}\ }\textbf {\bibinfo {volume} {95}},\ \bibinfo {pages}
  {226801} (\bibinfo {year} {2005}{\natexlab{a}})}\BibitemShut {NoStop}%
\bibitem [{\citenamefont {Kane}\ and\ \citenamefont
  {Mele}(2005{\natexlab{b}})}]{kane2005z}%
  \BibitemOpen
  \bibfield  {author} {\bibinfo {author} {\bibfnamefont {C.~L.}\ \bibnamefont
  {Kane}}\ and\ \bibinfo {author} {\bibfnamefont {E.~J.}\ \bibnamefont
  {Mele}},\ }\href@noop {} {\bibfield  {journal} {\bibinfo  {journal} {Physical
  Review Letters}\ }\textbf {\bibinfo {volume} {95}},\ \bibinfo {pages}
  {146802} (\bibinfo {year} {2005}{\natexlab{b}})}\BibitemShut {NoStop}%
\bibitem [{\citenamefont {Sichau}\ \emph {et~al.}(2019)\citenamefont {Sichau},
  \citenamefont {Prada}, \citenamefont {Anlauf}, \citenamefont {Lyon},
  \citenamefont {Bosnjak}, \citenamefont {Tiemann},\ and\ \citenamefont
  {Blick}}]{jonas}%
  \BibitemOpen
  \bibfield  {author} {\bibinfo {author} {\bibfnamefont {J.}~\bibnamefont
  {Sichau}}, \bibinfo {author} {\bibfnamefont {M.}~\bibnamefont {Prada}},
  \bibinfo {author} {\bibfnamefont {T.}~\bibnamefont {Anlauf}}, \bibinfo
  {author} {\bibfnamefont {T.~J.}\ \bibnamefont {Lyon}}, \bibinfo {author}
  {\bibfnamefont {B.}~\bibnamefont {Bosnjak}}, \bibinfo {author} {\bibfnamefont
  {L.}~\bibnamefont {Tiemann}}, \ and\ \bibinfo {author} {\bibfnamefont
  {R.}~\bibnamefont {Blick}},\ }\href@noop {} {\bibfield  {journal} {\bibinfo
  {journal} {Physical Review Letters}\ }\textbf {\bibinfo {volume} {122}},\
  \bibinfo {pages} {046403} (\bibinfo {year} {2019})}\BibitemShut {NoStop}%
\bibitem [{\citenamefont {Mecklenburg}\ and\ \citenamefont
  {Regan}(2011)}]{regan}%
  \BibitemOpen
  \bibfield  {author} {\bibinfo {author} {\bibfnamefont {M.}~\bibnamefont
  {Mecklenburg}}\ and\ \bibinfo {author} {\bibfnamefont {B.~C.}\ \bibnamefont
  {Regan}},\ }\href {\doibase 10.1103/PhysRevLett.106.116803} {\bibfield
  {journal} {\bibinfo  {journal} {Phys. Rev. Lett.}\ }\textbf {\bibinfo
  {volume} {106}},\ \bibinfo {pages} {116803} (\bibinfo {year}
  {2011})}\BibitemShut {NoStop}%
\bibitem [{\citenamefont {Soodchomshom}(2013)}]{chino}%
  \BibitemOpen
  \bibfield  {author} {\bibinfo {author} {\bibfnamefont {B.}~\bibnamefont
  {Soodchomshom}},\ }\href {\doibase 10.1088/0256-307x/30/12/126201} {\bibfield
   {journal} {\bibinfo  {journal} {Chinese Physics Letters}\ }\textbf {\bibinfo
  {volume} {30}},\ \bibinfo {pages} {126201} (\bibinfo {year}
  {2013})}\BibitemShut {NoStop}%
\bibitem [{\citenamefont {Song}\ \emph {et~al.}(2015)\citenamefont {Song},
  \citenamefont {Paltoglou}, \citenamefont {Liu}, \citenamefont {Zhu},
  \citenamefont {Gallardo}, \citenamefont {Tang}, \citenamefont {Xu},
  \citenamefont {Ablowitz}, \citenamefont {Efremidis},\ and\ \citenamefont
  {Chen}}]{Song15}%
  \BibitemOpen
  \bibfield  {author} {\bibinfo {author} {\bibfnamefont {D.}~\bibnamefont
  {Song}}, \bibinfo {author} {\bibfnamefont {V.}~\bibnamefont {Paltoglou}},
  \bibinfo {author} {\bibfnamefont {S.}~\bibnamefont {Liu}}, \bibinfo {author}
  {\bibfnamefont {Y.}~\bibnamefont {Zhu}}, \bibinfo {author} {\bibfnamefont
  {D.}~\bibnamefont {Gallardo}}, \bibinfo {author} {\bibfnamefont
  {L.}~\bibnamefont {Tang}}, \bibinfo {author} {\bibfnamefont {J.}~\bibnamefont
  {Xu}}, \bibinfo {author} {\bibfnamefont {M.}~\bibnamefont {Ablowitz}},
  \bibinfo {author} {\bibfnamefont {N.~K.}\ \bibnamefont {Efremidis}}, \ and\
  \bibinfo {author} {\bibfnamefont {Z.}~\bibnamefont {Chen}},\ }\href {\doibase
  10.1038/ncomms7272} {\bibfield  {journal} {\bibinfo  {journal} {Nature
  Comms.}\ }\textbf {\bibinfo {volume} {6}},\ \bibinfo {pages} {6272} (\bibinfo
  {year} {2015})}\BibitemShut {NoStop}%
\bibitem [{\citenamefont {Liu}\ \emph {et~al.}(2018)\citenamefont {Liu},
  \citenamefont {Song}, \citenamefont {Xia}, \citenamefont {Dai}, \citenamefont
  {Tang}, \citenamefont {Xu},\ and\ \citenamefont {Chen}}]{Liu2018}%
  \BibitemOpen
  \bibfield  {author} {\bibinfo {author} {\bibfnamefont {X.}~\bibnamefont
  {Liu}}, \bibinfo {author} {\bibfnamefont {D.}~\bibnamefont {Song}}, \bibinfo
  {author} {\bibfnamefont {S.}~\bibnamefont {Xia}}, \bibinfo {author}
  {\bibfnamefont {Z.}~\bibnamefont {Dai}}, \bibinfo {author} {\bibfnamefont
  {L.}~\bibnamefont {Tang}}, \bibinfo {author} {\bibfnamefont {J.}~\bibnamefont
  {Xu}}, \ and\ \bibinfo {author} {\bibfnamefont {Z.}~\bibnamefont {Chen}},\
  }in\ \href {\doibase 10.1364/CLEO_QELS.2018.FTh3E.4} {\emph {\bibinfo
  {booktitle} {Conference on Lasers and Electro-Optics}}}\ (\bibinfo
  {publisher} {Optical Society of America},\ \bibinfo {year} {2018})\ p.\
  \bibinfo {pages} {FTh3E.4}\BibitemShut {NoStop}%
\bibitem [{\citenamefont {Liu}\ \emph {et~al.}(2020)\citenamefont {Liu},
  \citenamefont {Xia}, \citenamefont {Jajtić}, \citenamefont {Song},
  \citenamefont {Li}, \citenamefont {Tang}, \citenamefont {Leykam},
  \citenamefont {Xu}, \citenamefont {Buljan},\ and\ \citenamefont
  {Chen}}]{Liu2020}%
  \BibitemOpen
  \bibfield  {author} {\bibinfo {author} {\bibfnamefont {X.}~\bibnamefont
  {Liu}}, \bibinfo {author} {\bibfnamefont {S.}~\bibnamefont {Xia}}, \bibinfo
  {author} {\bibfnamefont {E.}~\bibnamefont {Jajtić}}, \bibinfo {author}
  {\bibfnamefont {D.}~\bibnamefont {Song}}, \bibinfo {author} {\bibfnamefont
  {D.}~\bibnamefont {Li}}, \bibinfo {author} {\bibfnamefont {L.}~\bibnamefont
  {Tang}}, \bibinfo {author} {\bibfnamefont {D.}~\bibnamefont {Leykam}},
  \bibinfo {author} {\bibfnamefont {J.}~\bibnamefont {Xu}}, \bibinfo {author}
  {\bibfnamefont {H.}~\bibnamefont {Buljan}}, \ and\ \bibinfo {author}
  {\bibfnamefont {Z.}~\bibnamefont {Chen}},\ }\href {\doibase
  https://doi.org/10.1038/s41467-020-15374-x} {\bibfield  {journal} {\bibinfo
  {journal} {Nature Comms.}\ }\textbf {\bibinfo {volume} {11}},\ \bibinfo
  {pages} {1586} (\bibinfo {year} {2020})}\BibitemShut {NoStop}%
\bibitem [{\citenamefont {Prada}\ \emph {et~al.}(2020)\citenamefont {Prada},
  \citenamefont {Sichau}, \citenamefont {Tiemann},\ and\ \citenamefont
  {Blick}}]{exp}%
  \BibitemOpen
  \bibfield  {author} {\bibinfo {author} {\bibfnamefont {M.}~\bibnamefont
  {Prada}}, \bibinfo {author} {\bibfnamefont {J.}~\bibnamefont {Sichau}},
  \bibinfo {author} {\bibfnamefont {L.}~\bibnamefont {Tiemann}}, \ and\
  \bibinfo {author} {\bibfnamefont {R.}~\bibnamefont {Blick}},\ }\href@noop {}
  {\bibfield  {journal} {\bibinfo  {journal} {unpublished}\ } (\bibinfo {year}
  {2020})}\BibitemShut {NoStop}%
\bibitem [{\citenamefont {Castro~Neto}\ \emph {et~al.}(2009)\citenamefont
  {Castro~Neto}, \citenamefont {Guinea}, \citenamefont {Peres}, \citenamefont
  {Novoselov},\ and\ \citenamefont {Geim}}]{neto2009electronic}%
  \BibitemOpen
  \bibfield  {author} {\bibinfo {author} {\bibfnamefont {A.~H.}\ \bibnamefont
  {Castro~Neto}}, \bibinfo {author} {\bibfnamefont {F.}~\bibnamefont {Guinea}},
  \bibinfo {author} {\bibfnamefont {N.~M.~R.}\ \bibnamefont {Peres}}, \bibinfo
  {author} {\bibfnamefont {K.~S.}\ \bibnamefont {Novoselov}}, \ and\ \bibinfo
  {author} {\bibfnamefont {A.~K.}\ \bibnamefont {Geim}},\ }\href@noop {}
  {\bibfield  {journal} {\bibinfo  {journal} {Reviews of Modern Physics}\
  }\textbf {\bibinfo {volume} {81}},\ \bibinfo {pages} {109} (\bibinfo {year}
  {2009})}\BibitemShut {NoStop}%
\bibitem [{\citenamefont {Katsnelson}(2012)}]{katsnelson2012graphene}%
  \BibitemOpen
  \bibfield  {author} {\bibinfo {author} {\bibfnamefont {M.~I.}\ \bibnamefont
  {Katsnelson}},\ }\href@noop {} {\emph {\bibinfo {title} {Graphene: Carbon in
  Two Dimensions}}}\ (\bibinfo  {publisher} {Cambridge university press},\
  \bibinfo {year} {2012})\BibitemShut {NoStop}%
\bibitem [{\citenamefont {Semenoff}(1984)}]{semenov}%
  \BibitemOpen
  \bibfield  {author} {\bibinfo {author} {\bibfnamefont {G.~W.}\ \bibnamefont
  {Semenoff}},\ }\href {\doibase 10.1103/PhysRevLett.53.2449} {\bibfield
  {journal} {\bibinfo  {journal} {Phys. Rev. Lett.}\ }\textbf {\bibinfo
  {volume} {53}},\ \bibinfo {pages} {2449} (\bibinfo {year}
  {1984})}\BibitemShut {NoStop}%
\bibitem [{\citenamefont {Shapere}\ and\ \citenamefont
  {Wilczek}(1989)}]{shapere}%
  \BibitemOpen
  \bibfield  {author} {\bibinfo {author} {\bibfnamefont {A.}~\bibnamefont
  {Shapere}}\ and\ \bibinfo {author} {\bibfnamefont {F.}~\bibnamefont
  {Wilczek}},\ }\href@noop {} {\emph {\bibinfo {title} {Geometrical Phases in
  Physics}}}\ (\bibinfo  {publisher} {Singapore},\ \bibinfo {year}
  {1989})\BibitemShut {NoStop}%
\bibitem [{\citenamefont {Ando}\ \emph {et~al.}(1998)\citenamefont {Ando},
  \citenamefont {Nakanishi},\ and\ \citenamefont {Saito}}]{ando1998}%
  \BibitemOpen
  \bibfield  {author} {\bibinfo {author} {\bibfnamefont {T.}~\bibnamefont
  {Ando}}, \bibinfo {author} {\bibfnamefont {T.}~\bibnamefont {Nakanishi}}, \
  and\ \bibinfo {author} {\bibfnamefont {R.}~\bibnamefont {Saito}},\ }\href
  {\doibase 10.1143/jpsj.67.2857} {\bibfield  {journal} {\bibinfo  {journal}
  {Journal of the Physical Society of Japan}\ }\textbf {\bibinfo {volume}
  {67}},\ \bibinfo {pages} {2857} (\bibinfo {year} {1998})}\BibitemShut
  {NoStop}%
\bibitem [{\citenamefont {McEuen}\ \emph {et~al.}(1999)\citenamefont {McEuen},
  \citenamefont {Bockrath}, \citenamefont {Cobden}, \citenamefont {Yoon},\ and\
  \citenamefont {Louie}}]{mceuen1999}%
  \BibitemOpen
  \bibfield  {author} {\bibinfo {author} {\bibfnamefont {P.~L.}\ \bibnamefont
  {McEuen}}, \bibinfo {author} {\bibfnamefont {M.}~\bibnamefont {Bockrath}},
  \bibinfo {author} {\bibfnamefont {D.~H.}\ \bibnamefont {Cobden}}, \bibinfo
  {author} {\bibfnamefont {Y.-G.}\ \bibnamefont {Yoon}}, \ and\ \bibinfo
  {author} {\bibfnamefont {S.~G.}\ \bibnamefont {Louie}},\ }\href {\doibase
  10.1103/PhysRevLett.83.5098} {\bibfield  {journal} {\bibinfo  {journal}
  {Phys. Rev. Lett.}\ }\textbf {\bibinfo {volume} {83}},\ \bibinfo {pages}
  {5098} (\bibinfo {year} {1999})}\BibitemShut {NoStop}%
\bibitem [{\citenamefont {Hasan}\ and\ \citenamefont
  {Kane}(2010)}]{hasan2010colloquium}%
  \BibitemOpen
  \bibfield  {author} {\bibinfo {author} {\bibfnamefont {M.~Z.}\ \bibnamefont
  {Hasan}}\ and\ \bibinfo {author} {\bibfnamefont {C.~L.}\ \bibnamefont
  {Kane}},\ }\href@noop {} {\bibfield  {journal} {\bibinfo  {journal} {Reviews
  of Modern Physics}\ }\textbf {\bibinfo {volume} {82}},\ \bibinfo {pages}
  {3045} (\bibinfo {year} {2010})}\BibitemShut {NoStop}%
\bibitem [{\citenamefont {Asb\'oth}\ \emph {et~al.}(2006)\citenamefont
  {Asb\'oth}, \citenamefont {Oroszl\'any},\ and\ \citenamefont
  {P\'alyi}}]{asboth}%
  \BibitemOpen
  \bibfield  {author} {\bibinfo {author} {\bibfnamefont {J.~K.}\ \bibnamefont
  {Asb\'oth}}, \bibinfo {author} {\bibfnamefont {L.}~\bibnamefont
  {Oroszl\'any}}, \ and\ \bibinfo {author} {\bibfnamefont {A.}~\bibnamefont
  {P\'alyi}},\ }\href@noop {} {\emph {\bibinfo {title} {A Short Course on
  Topological Insulators}}}\ (\bibinfo  {publisher} {Springer},\ \bibinfo
  {year} {2006})\BibitemShut {NoStop}%
\bibitem [{\citenamefont {Prada}\ \emph {et~al.}(2011)\citenamefont {Prada},
  \citenamefont {San-Jose}, \citenamefont {Brey},\ and\ \citenamefont
  {Fertig}}]{eprada}%
  \BibitemOpen
  \bibfield  {author} {\bibinfo {author} {\bibfnamefont {E.}~\bibnamefont
  {Prada}}, \bibinfo {author} {\bibfnamefont {P.}~\bibnamefont {San-Jose}},
  \bibinfo {author} {\bibfnamefont {L.}~\bibnamefont {Brey}}, \ and\ \bibinfo
  {author} {\bibfnamefont {H.}~\bibnamefont {Fertig}},\ }\href {\doibase
  https://doi.org/10.1016/j.ssc.2011.05.016} {\bibfield  {journal} {\bibinfo
  {journal} {Solid State Communications}\ }\textbf {\bibinfo {volume} {151}},\
  \bibinfo {pages} {1075} (\bibinfo {year} {2011})}\BibitemShut {NoStop}%
\bibitem [{\citenamefont {Konschuh}\ \emph {et~al.}(2010)\citenamefont
  {Konschuh}, \citenamefont {Gmitra},\ and\ \citenamefont
  {Fabian}}]{konschuh2010tight}%
  \BibitemOpen
  \bibfield  {author} {\bibinfo {author} {\bibfnamefont {S.}~\bibnamefont
  {Konschuh}}, \bibinfo {author} {\bibfnamefont {M.}~\bibnamefont {Gmitra}}, \
  and\ \bibinfo {author} {\bibfnamefont {J.}~\bibnamefont {Fabian}},\
  }\href@noop {} {\bibfield  {journal} {\bibinfo  {journal} {Physical Review
  B}\ }\textbf {\bibinfo {volume} {82}},\ \bibinfo {pages} {245412} (\bibinfo
  {year} {2010})}\BibitemShut {NoStop}%
\bibitem [{\citenamefont {Huertas-Hernando}\ \emph {et~al.}(2006)\citenamefont
  {Huertas-Hernando}, \citenamefont {Guinea},\ and\ \citenamefont
  {Brataas}}]{huertas2006spin}%
  \BibitemOpen
  \bibfield  {author} {\bibinfo {author} {\bibfnamefont {D.}~\bibnamefont
  {Huertas-Hernando}}, \bibinfo {author} {\bibfnamefont {F.}~\bibnamefont
  {Guinea}}, \ and\ \bibinfo {author} {\bibfnamefont {A.}~\bibnamefont
  {Brataas}},\ }\href@noop {} {\bibfield  {journal} {\bibinfo  {journal}
  {Physical Review B}\ }\textbf {\bibinfo {volume} {74}},\ \bibinfo {pages}
  {155426} (\bibinfo {year} {2006})}\BibitemShut {NoStop}%
\bibitem [{\citenamefont {Min}\ \emph {et~al.}(2006)\citenamefont {Min},
  \citenamefont {Hill}, \citenamefont {Sinitsyn}, \citenamefont {Sahu},
  \citenamefont {Kleinman},\ and\ \citenamefont
  {MacDonald}}]{min2006intrinsic}%
  \BibitemOpen
  \bibfield  {author} {\bibinfo {author} {\bibfnamefont {H.}~\bibnamefont
  {Min}}, \bibinfo {author} {\bibfnamefont {J.~E.}\ \bibnamefont {Hill}},
  \bibinfo {author} {\bibfnamefont {N.~A.}\ \bibnamefont {Sinitsyn}}, \bibinfo
  {author} {\bibfnamefont {B.~R.}\ \bibnamefont {Sahu}}, \bibinfo {author}
  {\bibfnamefont {L.}~\bibnamefont {Kleinman}}, \ and\ \bibinfo {author}
  {\bibfnamefont {A.~H.}\ \bibnamefont {MacDonald}},\ }\href@noop {} {\bibfield
   {journal} {\bibinfo  {journal} {Physical Review B}\ }\textbf {\bibinfo
  {volume} {74}},\ \bibinfo {pages} {165310} (\bibinfo {year}
  {2006})}\BibitemShut {NoStop}%
\bibitem [{\citenamefont {Yao}\ \emph {et~al.}(2007)\citenamefont {Yao},
  \citenamefont {Ye}, \citenamefont {Qi}, \citenamefont {Zhang},\ and\
  \citenamefont {Fang}}]{yao2007spin}%
  \BibitemOpen
  \bibfield  {author} {\bibinfo {author} {\bibfnamefont {Y.}~\bibnamefont
  {Yao}}, \bibinfo {author} {\bibfnamefont {F.}~\bibnamefont {Ye}}, \bibinfo
  {author} {\bibfnamefont {X.-L.}\ \bibnamefont {Qi}}, \bibinfo {author}
  {\bibfnamefont {S.-C.}\ \bibnamefont {Zhang}}, \ and\ \bibinfo {author}
  {\bibfnamefont {Z.}~\bibnamefont {Fang}},\ }\href@noop {} {\bibfield
  {journal} {\bibinfo  {journal} {Physical Review B}\ }\textbf {\bibinfo
  {volume} {75}},\ \bibinfo {pages} {041401} (\bibinfo {year}
  {2007})}\BibitemShut {NoStop}%
\bibitem [{\citenamefont {Rashba}(2009)}]{rashba2009graphene}%
  \BibitemOpen
  \bibfield  {author} {\bibinfo {author} {\bibfnamefont {E.~I.}\ \bibnamefont
  {Rashba}},\ }\href@noop {} {\bibfield  {journal} {\bibinfo  {journal}
  {Physical Review B}\ }\textbf {\bibinfo {volume} {79}},\ \bibinfo {pages}
  {161409} (\bibinfo {year} {2009})}\BibitemShut {NoStop}%
\bibitem [{\citenamefont {Gmitra}\ \emph {et~al.}(2009)\citenamefont {Gmitra},
  \citenamefont {Konschuh}, \citenamefont {Ertler}, \citenamefont
  {Ambrosch-Draxl},\ and\ \citenamefont {Fabian}}]{gmitra2009band}%
  \BibitemOpen
  \bibfield  {author} {\bibinfo {author} {\bibfnamefont {M.}~\bibnamefont
  {Gmitra}}, \bibinfo {author} {\bibfnamefont {S.}~\bibnamefont {Konschuh}},
  \bibinfo {author} {\bibfnamefont {C.}~\bibnamefont {Ertler}}, \bibinfo
  {author} {\bibfnamefont {C.}~\bibnamefont {Ambrosch-Draxl}}, \ and\ \bibinfo
  {author} {\bibfnamefont {J.}~\bibnamefont {Fabian}},\ }\href@noop {}
  {\bibfield  {journal} {\bibinfo  {journal} {Physical Review B}\ }\textbf
  {\bibinfo {volume} {80}},\ \bibinfo {pages} {235431} (\bibinfo {year}
  {2009})}\BibitemShut {NoStop}%
\bibitem [{\citenamefont {Kochan}\ \emph {et~al.}(2017)\citenamefont {Kochan},
  \citenamefont {Irmer},\ and\ \citenamefont {Fabian}}]{kochan}%
  \BibitemOpen
  \bibfield  {author} {\bibinfo {author} {\bibfnamefont {D.}~\bibnamefont
  {Kochan}}, \bibinfo {author} {\bibfnamefont {S.}~\bibnamefont {Irmer}}, \
  and\ \bibinfo {author} {\bibfnamefont {J.}~\bibnamefont {Fabian}},\ }\href
  {\doibase 10.1103/PhysRevB.95.165415} {\bibfield  {journal} {\bibinfo
  {journal} {Phys. Rev. B}\ }\textbf {\bibinfo {volume} {95}},\ \bibinfo
  {pages} {165415} (\bibinfo {year} {2017})}\BibitemShut {NoStop}%
\bibitem [{\citenamefont {Robinson}\ \emph {et~al.}(2008)\citenamefont
  {Robinson}, \citenamefont {Schomerus}, \citenamefont {Oroszl\'any},\ and\
  \citenamefont {Fal'ko}}]{falko}%
  \BibitemOpen
  \bibfield  {author} {\bibinfo {author} {\bibfnamefont {J.~P.}\ \bibnamefont
  {Robinson}}, \bibinfo {author} {\bibfnamefont {H.}~\bibnamefont {Schomerus}},
  \bibinfo {author} {\bibfnamefont {L.}~\bibnamefont {Oroszl\'any}}, \ and\
  \bibinfo {author} {\bibfnamefont {V.~I.}\ \bibnamefont {Fal'ko}},\ }\href
  {\doibase 10.1103/PhysRevLett.101.196803} {\bibfield  {journal} {\bibinfo
  {journal} {Phys. Rev. Lett.}\ }\textbf {\bibinfo {volume} {101}},\ \bibinfo
  {pages} {196803} (\bibinfo {year} {2008})}\BibitemShut {NoStop}%
\bibitem [{\citenamefont {Liu}\ \emph {et~al.}(2011)\citenamefont {Liu},
  \citenamefont {Jiang},\ and\ \citenamefont {Yao}}]{chengLiu}%
  \BibitemOpen
  \bibfield  {author} {\bibinfo {author} {\bibfnamefont {C.-C.}\ \bibnamefont
  {Liu}}, \bibinfo {author} {\bibfnamefont {H.}~\bibnamefont {Jiang}}, \ and\
  \bibinfo {author} {\bibfnamefont {Y.}~\bibnamefont {Yao}},\ }\href {\doibase
  10.1103/PhysRevB.84.195430} {\bibfield  {journal} {\bibinfo  {journal} {Phys.
  Rev. B}\ }\textbf {\bibinfo {volume} {84}},\ \bibinfo {pages} {195430}
  (\bibinfo {year} {2011})}\BibitemShut {NoStop}%
\bibitem [{\citenamefont {Gmitra}\ \emph {et~al.}(2013)\citenamefont {Gmitra},
  \citenamefont {Kochan},\ and\ \citenamefont {Fabian}}]{kochan2}%
  \BibitemOpen
  \bibfield  {author} {\bibinfo {author} {\bibfnamefont {M.}~\bibnamefont
  {Gmitra}}, \bibinfo {author} {\bibfnamefont {D.}~\bibnamefont {Kochan}}, \
  and\ \bibinfo {author} {\bibfnamefont {J.}~\bibnamefont {Fabian}},\ }\href
  {\doibase 10.1103/PhysRevLett.110.246602} {\bibfield  {journal} {\bibinfo
  {journal} {Phys. Rev. Lett.}\ }\textbf {\bibinfo {volume} {110}},\ \bibinfo
  {pages} {246602} (\bibinfo {year} {2013})}\BibitemShut {NoStop}%
\bibitem [{\citenamefont {Banszerus}\ \emph {et~al.}(2020)\citenamefont
  {Banszerus}, \citenamefont {Frohn}, \citenamefont {Fabian}, \citenamefont
  {Somanchi}, \citenamefont {Epping}, \citenamefont {M\"uller}, \citenamefont
  {Neumaier}, \citenamefont {Watanabe}, \citenamefont {Taniguchi},
  \citenamefont {Libisch}, \citenamefont {Beschoten}, \citenamefont {Hassler},\
  and\ \citenamefont {Stampfer}}]{BanszerusPRL20}%
  \BibitemOpen
  \bibfield  {author} {\bibinfo {author} {\bibfnamefont {L.}~\bibnamefont
  {Banszerus}}, \bibinfo {author} {\bibfnamefont {B.}~\bibnamefont {Frohn}},
  \bibinfo {author} {\bibfnamefont {T.}~\bibnamefont {Fabian}}, \bibinfo
  {author} {\bibfnamefont {S.}~\bibnamefont {Somanchi}}, \bibinfo {author}
  {\bibfnamefont {A.}~\bibnamefont {Epping}}, \bibinfo {author} {\bibfnamefont
  {M.}~\bibnamefont {M\"uller}}, \bibinfo {author} {\bibfnamefont
  {D.}~\bibnamefont {Neumaier}}, \bibinfo {author} {\bibfnamefont
  {K.}~\bibnamefont {Watanabe}}, \bibinfo {author} {\bibfnamefont
  {T.}~\bibnamefont {Taniguchi}}, \bibinfo {author} {\bibfnamefont
  {F.}~\bibnamefont {Libisch}}, \bibinfo {author} {\bibfnamefont
  {B.}~\bibnamefont {Beschoten}}, \bibinfo {author} {\bibfnamefont
  {F.}~\bibnamefont {Hassler}}, \ and\ \bibinfo {author} {\bibfnamefont
  {C.}~\bibnamefont {Stampfer}},\ }\href {\doibase
  10.1103/PhysRevLett.124.177701} {\bibfield  {journal} {\bibinfo  {journal}
  {Phys. Rev. Lett.}\ }\textbf {\bibinfo {volume} {124}},\ \bibinfo {pages}
  {177701} (\bibinfo {year} {2020})}\BibitemShut {NoStop}%
\bibitem [{\citenamefont {McCann}\ and\ \citenamefont
  {Fal'ko}(2006)}]{mccannBLG}%
  \BibitemOpen
  \bibfield  {author} {\bibinfo {author} {\bibfnamefont {E.}~\bibnamefont
  {McCann}}\ and\ \bibinfo {author} {\bibfnamefont {V.~I.}\ \bibnamefont
  {Fal'ko}},\ }\href {\doibase 10.1103/PhysRevLett.96.086805} {\bibfield
  {journal} {\bibinfo  {journal} {Phys. Rev. Lett.}\ }\textbf {\bibinfo
  {volume} {96}},\ \bibinfo {pages} {086805} (\bibinfo {year}
  {2006})}\BibitemShut {NoStop}%
\bibitem [{\citenamefont {Novoselov}\ \emph {et~al.}(2006)\citenamefont
  {Novoselov}, \citenamefont {McCann}, \citenamefont {Morozov}, \citenamefont
  {Fal'ko}, \citenamefont {Katsnelson}, \citenamefont {Zeitler}, \citenamefont
  {Jiang}, \citenamefont {Schedin},\ and\ \citenamefont
  {Geim}}]{Novoselov2006}%
  \BibitemOpen
  \bibfield  {author} {\bibinfo {author} {\bibfnamefont {K.~S.}\ \bibnamefont
  {Novoselov}}, \bibinfo {author} {\bibfnamefont {E.}~\bibnamefont {McCann}},
  \bibinfo {author} {\bibfnamefont {S.~V.}\ \bibnamefont {Morozov}}, \bibinfo
  {author} {\bibfnamefont {V.~I.}\ \bibnamefont {Fal'ko}}, \bibinfo {author}
  {\bibfnamefont {M.~I.}\ \bibnamefont {Katsnelson}}, \bibinfo {author}
  {\bibfnamefont {U.}~\bibnamefont {Zeitler}}, \bibinfo {author} {\bibfnamefont
  {D.}~\bibnamefont {Jiang}}, \bibinfo {author} {\bibfnamefont
  {F.}~\bibnamefont {Schedin}}, \ and\ \bibinfo {author} {\bibfnamefont
  {A.~K.}\ \bibnamefont {Geim}},\ }\href {\doibase 10.1038/nphys245} {\bibfield
   {journal} {\bibinfo  {journal} {Nature Physics}\ }\textbf {\bibinfo {volume}
  {2}},\ \bibinfo {pages} {177} (\bibinfo {year} {2006})}\BibitemShut {NoStop}%
\bibitem [{\citenamefont {Slater}\ and\ \citenamefont {Koster}(1954)}]{slater}%
  \BibitemOpen
  \bibfield  {author} {\bibinfo {author} {\bibfnamefont {J.~C.}\ \bibnamefont
  {Slater}}\ and\ \bibinfo {author} {\bibfnamefont {G.~F.}\ \bibnamefont
  {Koster}},\ }\href {\doibase 10.1103/PhysRev.94.1498} {\bibfield  {journal}
  {\bibinfo  {journal} {Phys. Rev.}\ }\textbf {\bibinfo {volume} {94}},\
  \bibinfo {pages} {1498} (\bibinfo {year} {1954})}\BibitemShut {NoStop}%
\end{thebibliography}%

\end{document}